\documentclass[useAMS,usenatbib]{mn2e}
\usepackage{times}
\usepackage{bigints}
\usepackage[pdftex]{graphicx}
\usepackage{amsmath}
\usepackage{amssymb}
\usepackage{natbib}
\usepackage{hyperref} 

\usepackage{newlfont,verbatim,enumerate,array}
\usepackage[usenames,dvipsnames,svgnames,table]{xcolor}
\newcommand\vect[1]{\ensuremath{\mathbf{#1}}}

\newcommand\velocity{\ensuremath{\text{km\,s}^{-1}}}
\newcommand\patspeed{\ensuremath{\text{km\,s}^{-1}\,\text{kpc}^{-1}}}

\newcommand\amuse{\textsc{Amuse}}
\newcommand\bridge{\textsc{Bridge}}

\newcommand\tenc{\ensuremath{t_\mathrm{enc}}}
\newcommand\renc{\ensuremath{r_\mathrm{enc}}}
\newcommand\venc{\ensuremath{v_\mathrm{enc}}}
\newcommand\menc{\ensuremath{M_\mathrm{enc}}}
\newcommand\enc{\ensuremath{_\mathrm{enc}}}
\usepackage{xcolor}

\title[Stellar encounters along a migrating orbit of the Sun]{The rate of stellar encounters along a migrating  orbit of the Sun}
\author[C.A. Mart\'inez-Barbosa et al.]{C.A. Mart\'inez-Barbosa,\thanks{E-mail:\,\,cmartinez@strw.leidenuniv.nl\,\,(CAMB), jilkova@strw.leidenuniv.nl\,\,(LJ)}$\dag$ 
L. J\'ilkov\'a, $^{\star}$\thanks{Both authors contributed equally to this work.}
S. Portegies Zwart, 
and A.G.A. Brown\\
Leiden Observatory, University of Leiden, P.B. 9513, Leiden 2300 RA, the Netherlands }
\begin{document}

\date{Accepted XXXXXXXXXXX. Received XXXXXXXXXXXX; in original form XXXXXXXXXXXXX}

\pagerange{\pageref{firstpage}--\pageref{lastpage}} \pubyear{2002}

\maketitle

\label{firstpage}

\begin{abstract}
The frequency of Galactic stellar encounters the Solar system experienced depends on the local density and velocity dispersion along the orbit of the Sun in the Milky Way galaxy.
We aim at determining the effect of the radial migration of the solar orbit on the rate of stellar encounters.  As a first step we integrate the orbit of the Sun backwards in time in an analytical potential of the Milky Way. We use the present-day phase-space coordinates of the Sun, according to the measured uncertainties. The resulting orbits are inserted in an N-body simulation of the Galaxy, where the stellar velocity dispersion is calculated at each position along the orbit of the Sun.
We compute the rate of Galactic stellar encounters by employing three different solar orbits \,---\,migrating from the inner disk, without any substantial migration, and migrating from the outer disk.
We find that the rate for encounters within $4\times10^5$\,AU from the Sun is about 21, 39 and 63\,Myr$^{-1}$, respectively.
The stronger encounters establish the outer limit of the so-called parking zone, which is the region in the plane of the orbital eccentricities and semi-major axes where the planetesimals of the Solar system have been perturbed only by interactions with stars belonging to the Sun's birth cluster.
We estimate the outer edge of the parking zone at semi-major axes of  250--1300\,AU (the outward and inward migrating orbits reaching the smallest and largest values, respectively), which is one order of magnitude smaller than the  determination made by \cite{portegies15}.
We further discuss the effect of stellar encounters on the stability of the hypothetical Planet~9.
\end{abstract}


\section{Introduction}\label{Sect:introd3}

To explain the constant rate of observed new long period comets, \citet{oort} suggested that more than $10^{11}$ icy bodies orbit the Sun with aphelia of $5$--$15\times10^4$~AU, and isotropically distributed inclinations of their orbital planes. 
The comets are delivered to the inner Solar system from the cloud due to perturbation by the Galactic tide and passing stars (see for example \citealp{rickman14} or \citealp{2015SSRv..197..191D} for summaries), and the interstellar medium such as the giant molecular clouds \citep[e.g.][]{1985AJ.....90.1548H,1996A&A...308..988B,2009CoSka..39...85J,2008CoSka..38...33J}.

The Galactic tide has a stronger overall effect when averaged over long time scales \citep[for example][]{heisler86}.
The effect of the encounters is stochastic and helps to keep the Oort cloud isotropic \citep[e.g.][and references therein]{kaib11}.
The two mechanisms act together and combine in a non-linear way \citep{rickman08,2011Icar..214..334F}.

The orbit of the Sun in the Galaxy determines the intensity of the gravitational tides the Solar
System was exposed to, as well as the number of stars around the Sun that could pass close enough to
perturb the Oort cloud. \citet{kaib11} investigated the effect of encounters with the field stars
and that of the Galactic tides on the Oort cloud, considering the so-called radial migration effect on the orbit of the Sun \citep[see e.g][for a more detailed description]{sellwood,roskar, minchev10, martinezb14}. 
They simulated the Oort cloud around the Sun, adopting possible solar orbits from the simulation of a
Milky Way-like galaxy of \citet{roskar}, including those that experienced no migration and those
that experienced strong radial migration (some of their solar analogues get as close as $2$~kpc from
the Galactic centre or as far as $13$~kpc). Kaib and collaborators found that the present-day structure
of the Oort Cloud strongly depends on the Sun's orbital history, in particular on its minimum
past Galactocentric distance. The inner edge of the Oort cloud shows a similar
dependence (on the orbital history of the Sun) and it is also influenced by the effect of
 strong encounters between the Sun and other stars.


With the increasing amount of precise astrometric and radial velocity data for the stars in the
solar neighborhood, several studies have focused on the identification of stars that passed close to
the Solar system in the recent past, or will pass close by in the future \citep[][]{mamajek,
bailer15, dubinsky15}. \citet{mamajek} identified the star that is currently known to
have made the closest approach to the Sun --- the so called Scholz's star that passed the Solar
System at $0.25^{+0.11}_{-0.07}$\,pc. Additionally, \citet{2015MNRAS.454.3267F} studied the effect of recent and future
stellar encounters on the flux of the long period comets.  They carried out simulations of the Oort cloud, considering
perturbations by the identified encounters and a constant Galactic field at the current solar
Galactocentric radius, and kept track of the flux of long-period comets injected into the inner
Solar system as a consequence of the encounters. Unlike \citet{kaib11}, \citet{2015MNRAS.454.3267F} focused only
on the effect of the actually observed perturbers.  They conclude that past encounters in their
sample explain about 5\% of the currently observed long period comets and they suggest that the
Solar system experienced more strong, as yet unidentified, encounters. 

\cite{portegies15} discuss the effect of the stellar encounter history on the structure of the
system of planetesimals surrounding the Sun. They considered encounters with stars in the Sun's
birth cluster (early on in the history of the Sun) and encounters with field stars that occur as
the Sun orbits in the Galaxy. The encounters with the field stars set the outer edge of the so called
Parking zone of the Solar system \citep{portegies15}. The parking zone is defined as a region in the
plane of semi-major axis and eccentricity of objects orbiting the Sun that have been perturbed by the
parental star cluster but not by the planets or the Galactic perturbations. The orbits located in
the parking zone maintain a record of the interaction of the Solar system with stars belonging to
the Sun's birth cluster. Therefore, these orbits carry information that can constrain the natal
environment of the Sun. Recently, \citet{jilkova15} argued that a population of observed
planetesimals with semi-major axes $>150$\,au and perihelia $>30$\,au, would live in the parking zone
of the Solar system. They also found that such a  population might have been captured from a debris
disc of another star during a close flyby that happened in the Sun's birth cluster.

The outer edge of the parking zone is defined by the strongest encounter the Solar system
experienced after it left its birth cluster. The strength of the encounter is measured by the
perturbation of semi-major axes and eccentricity of the bodies in their orbit around the Sun.
\citet{portegies15} used the impulse approximation \citep{rickman76} to estimate the effect and
defined the outer edge of the Solar system's parking zone as corresponding to the perturbation
caused by the Scholz's star \citep{mamajek}. However, stronger encounters might have happened in the past,
as the Sun orbited in the Galactic disc. These encounters would alter the outer edge of the Solar system's parking zone moving it closer to the Sun. The perturbation strength of the stellar encounters depends on the characteristics of
the close encounters with field stars --- the mass of the other star, its closest approach and
relative velocity.  Similar to Scholz's star, the parameters of some of the recent close encounters
can be derived from the observed data \citep[for example][]{2015MNRAS.454.3267F, dubinsky15}.

Estimates of the number and strength of past encounters are difficult to make because of the large
uncertainties in the Galactic environment where the Sun has been moving since it left its birth cluster.  These uncertainties are due to the unknown evolution of the Galactic potential (leading to  uncertainties in the Sun's past orbit), which is in turn related to the unknown (population dependent) density and velocity dispersion of the Milky Way stars along the Sun's orbit. \cite{garcia01} studied the recent
encounter history of the Sun by integrating its orbit in an analytical Milky Way potential together
with 595 stars from the Hipparcos catalogue in order to identify recent and near future encounters.
In addition they estimated the encounter frequency for the Sun in its present environment by
considering the velocity dispersions and number densities of different types of stars.
\cite{rickman08} simulated the stellar encounters by assuming random encounter times (for a fixed
number of encounters) over 5 billion years and using velocity dispersions for 13 different types of
stars (different masses), with relative encounter frequencies for these types taken from
\cite{garcia01}. An alternative approach based on a numerical model of the Milky Way
was taken by \cite{kaib11}. The orbits of solar analogues in this model were extracted  from a simulation of a Milky Way-like galaxy and then the
encounters were simulated by tracking the stellar number density and velocity dispersion along the
orbit and then generating random encounters by starting stars at random orientations $1$~pc from the
Sun. The encounter velocities were generated using the recipe by \cite{rickman08}.

In this paper we aim to improve the determination of the  outer edge of the Solar system's parking zone by determining the number of stellar encounters experienced by the Sun along its orbit.  We compute the number of encounters  by employing the largest  Milky Way
simulation to date, which contains  51 billion particles, divided over a central bulge, a disk and a dark matter halo \citep[][]{bedorf}. This Galaxy model is used to estimate the velocity dispersion of the stars encountered by
the Sun along its orbit. To achieve this we integrate the Sun's orbit back in time using an
analytical potential for the Milky Way. The orbit of the Sun is then inserted in a snapshot of the  particle
simulation and the velocity dispersion of the disk stars is estimated at each position. We employ
three different orbits of the Sun (no radial migration, migration inward, migration outward) and
use the resulting estimates of the encounter frequencies along each of these orbits to discuss
the implications for the location of the outer edge of the Solar system's parking zone. We also discuss the effect of such encounters on the
stability of the orbit of the so-called Planet 9. 
 The presence of this object was predicted by \citet{BB16} in the outer Solar system to explain the clustering of the orbital elements of the distant Kuiper Belt Objects (KBOs).
According to the updated simulations of \cite{2016ApJ...824L..23B}, Planet 9 has a mass of 5--20\,M$_\oplus$; an eccentricity of $\sim 0.2$--0.8, semi-major axis of $\sim 500$--1050\,AU and perihelion distance of $\sim150$--350\,AU.

This paper is organized as follows: In Sect.\ \ref{sect:gal_model} we
explain the Galaxy model and we show three possible orbital histories
of the Sun. In Sect.\ \ref{sect:nenc} we determine the number of encounters along each of
these solar orbits. From this estimate, we generate a set of stellar encounters with random mass, encounter distance and velocity.  In Sect.\ \ref{sect:stability} we find the stellar encounters that produce the strongest perturbation of objects orbiting the Sun. These encounters are used to estimate the outer edge
of the Solar system's parking zone. In Sect. \ref{sect:discussion3} we discuss the effects of such encounters on the stability of the orbit of Planet 9. We also mention the limitations of our computations and the improvements that could be made in future studies. In Sect. \ref{sect:summary} we summarize.

\section{Galaxy model and possible orbital histories of the Sun}
\label{sect:gal_model}

\begin{table}
  \caption{Modeling parameters of the Milky Way.}
  \label{tab:params}
  \begin{tabular}{lll} \hline
    \multicolumn{3}{c}{ \rule{0pt}{3ex}\textit{Axisymmetric component}} \\ 
    \rule{0pt}{4ex}Mass of the bulge ($M_\mathrm{b}$) & $1.41\times 10^{10}$ M$_{\odot}$ &\\ 
    Scale length bulge ($b_\mathrm{1}$) & $0.3873$ kpc &\\
    Disk mass ($M_\mathrm{d}$) & $8.56\times10^{10}$ M$_{\odot}$  &\\
    Scale length 	1 disk ($a_\mathrm{2}$) & $5.31$ kpc  & 1) \\
    Scale length 2 disk ($b_\mathrm{2}$) & $0.25$ kpc   &\\
    Halo mass ($M_\mathrm{h}$) & $1.07\times 10^{11} $ M$_{\odot}$   &\\
    Scale length halo ($a_\mathrm{3}$) & 12 kpc  &\\ 
    \multicolumn{3}{c}{ \rule{0pt}{3ex}\textit{Central Bar}}   \\ 
     \rule{0pt}{4ex}Pattern speed ($\Omega_\mathrm{bar}$) & $55$ \patspeed & 2)\\ 
    Mass ($M_\mathrm{bar}$) & $9.8\times10^9$ M$_{\odot}$& 4) \\ 
    Semi-major axis ($a$) & $3.1$ kpc & 5)\\
    Axis ratio ($b/a$) & $0.37$& 5) \\
    Vertical axis ($c$) & 1 kpc & 6)\\
    Present-day orientation & $20^\circ$ & 3)\\ 
    \multicolumn{2}{c}{ \rule{0pt}{3ex}\textit{ Spiral arms}} \\ 
     \rule{0pt}{4ex}Pattern speed ($\Omega_\mathrm{sp}$) & $25$ \patspeed & 2)\\
    Number of spiral arms ($m$) & $2$ & 7)\\
    Amplitude ($A_\mathrm{sp}$) &  $3.9\times10^7$ M$_\odot$~kpc$^{-3}$ & 4) \\
    Pitch angle ($i$) & $ 15.5^\circ$ & 4)\\
    Scale length ($R_\mathrm{{\Sigma}}$) & $2.6$ kpc & 7)\\
    Scale height ($H$) & 0.3 kpc & 7)\\
    Present-day orientation & $20 ^\circ$ & 5) \\
    		        \hline
  \end{tabular}\\
  \textbf{References:} 1) \cite{allen}; 2) \cite{gerhard}; \\3) \cite{merce2}; 4) \cite{jilkova12}; \\ 5) \cite{martinezb14}; 6) \cite{monari14}; \\ 7) \cite{drimmel00}; 8) \cite{juric08}
\end{table}

We use an analytical potential to model the Milky Way. This potential is used to calculate possible solar orbits. The Galactic potential contains an axisymmetric and non-axisymmetric  components.  The axisymmetric part  contains a bulge, disk and dark matter
halo. The non-axisymmetric part contains a bar and spiral arms, which rotate as rigid bodies with different pattern speeds.

Given the configuration of the Galactic potential, we define three coordinate systems:
\begin{itemize} 
\item An inertial system that is fixed at the centre of the Galaxy, whose coordinates are denoted by  $\vect{x}=$ ($x$, $y$, $z$).
\item A system that corotates with the bar, whose coordinates are denoted by $\vect{x}_\mathrm{rot}=$ ($x_\mathrm{rot}$, $y_\mathrm{rot}$, $z_\mathrm{rot}$).  In this frame, the bar is located along the $x$-axis. The initial orientation and velocity of this rotating system correspond to the present-day orientation and pattern speed of the bar respectively (see Table \ref{tab:params}).
\item A system that corotates with the spiral arms, whose coordinates are denoted by $\vect{x}_\mathrm{rot_1}=$ ($x_\mathrm{rot_1}$, $y_\mathrm{rot_1}$, $z_\mathrm{rot_1}$). The initial orientation and velocity of this rotating system correspond to the present-day orientation and pattern speed of the spiral arms (see Table \ref{tab:params})
\end{itemize}

\begin{figure}
  \centering
  \includegraphics[width=84mm]{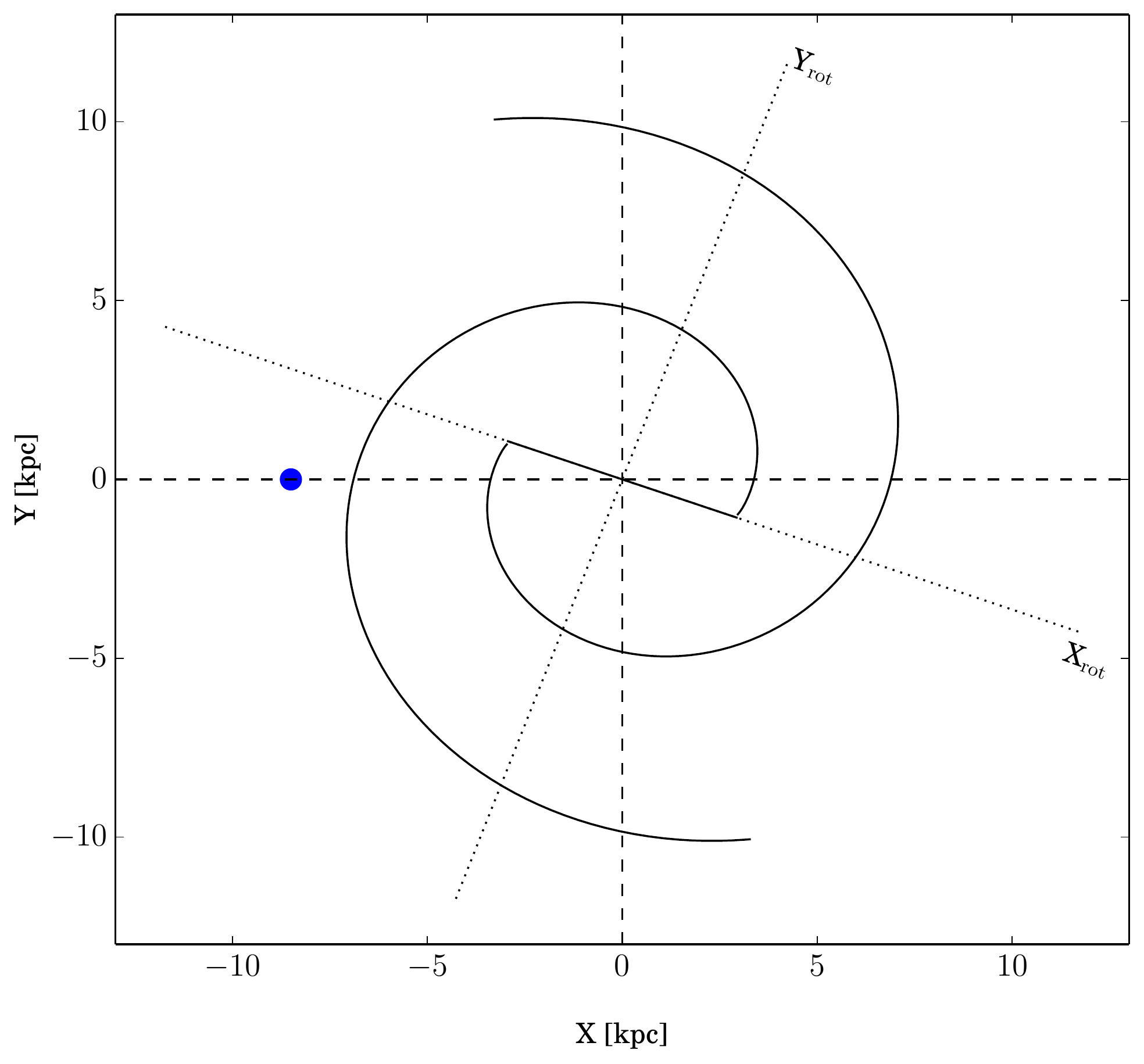}
  \caption{Configuration of the bar and spiral arms of the Galaxy at the present time. The blue circle marks the current  position of the Sun measured in an inertial system that is fixed at the center of the Galaxy. The axes X$_\mathrm{rot}$ and Y$_\mathrm{rot}$ correspond to a system that corotates with the bar. Note that the spiral arms start at the edges of the bar, therefore the coordinates (X$_\mathrm{rot_1}, $Y$_\mathrm{rot_1}$) and  (X$_\mathrm{rot}$, Y$_\mathrm{rot}$) overlap at the present time. \label{fig:pos_sun}}  
  \end{figure}

The reference systems explained above are shown in Fig. \ref{fig:pos_sun} and we use them to compute the components of the Galactic potential. The Axisymmetric potential is calculated in the inertial frame while the potential of the bar and spiral arms are calculated in their respective co-rotating frames. We however, compute the orbit of the Sun in the inertial system. Therefore, we use coordinate transformations to go from $\vect{x}_\mathrm{rot}$ or $\vect{x}_\mathrm{rot_1}$ to $\vect{x}$.

Hereafter the coordinates $r$ and $R$ represent the spherical and cylindrical radii. $\varphi$ is the angle measured from the $x$-axis  and in the direction opposite to the Galactic rotation (i.e. counterclockwise).  $z$ is the vertical component, perpendicular to the plane of the Galactic disk.

In Sects.\ \ref{sect:axi}- \ref{sect:sp} we give a detailed description of the axisymmetric and
non-axisymmetric components of the Galactic potential.

\subsection{Axisymmetric component}
\label{sect:axi}

As mentioned before, the axisymmetric component of the Galaxy consists of a bulge, disk and a dark
matter halo. We model the bulge of the Milky Way as a Plummer potential \citep{plummer}:

\begin{equation}
\Phi_\mathrm{bulge}= -\frac{GM_\mathrm{b}}{\sqrt{r^2 +b_1^2}}\,,
\label{eq:pbulge}
\end{equation}
where $G$ corresponds to the gravitational constant, $M_\mathrm{b}$ is the mass of the bulge, and
$b_1$ is its corresponding scale length. 

The disk of the Milky Way was modelled by using a Miyamoto-Nagai potential \citep{miyamoto}, which
is des\-cribed by the expression:
\begin{equation}
\Phi_\mathrm{disk}= -\frac{GM_\mathrm{d}}{\sqrt{R^2+ \left(a_2 +\sqrt{z^2+ b_2^2}\right)^2 }}\,.
\label{eq:pdisk}
\end{equation}
Here $M_\mathrm{d}$ corresponds to the mass of the disk. The parameters $a_2$ and $b_2$ are
constants that modulate its shape. In particular, when $a_2=0$, Eq.\ \ref{eq:pdisk} represents a
spherical distribution of mass. In the case where $b_2=0$, Eq.\ \ref{eq:pdisk} corresponds to the
potential of a completely flattened disk. 

Finally, we model the dark matter halo by means of a logarithmic potential of the form:

\begin{equation}
\begin{split}
\Phi_\mathrm{halo}= & -\frac{GM(r)}{r} \\ 
&- \frac{GM_\mathrm{h}}{1.02a_3}\left[-\frac{1.02}{1+\mathfrak{R}^{1.02}}
+\ln{\left(1+\mathfrak{R}^{1.02} \right)}  \right]_{r}^{100}\,,
\label{eq:phalo}
\end{split}
\end{equation}
where
\begin{align*}
M(r) &= \frac{M_\mathrm{h}\mathfrak{R}^{2.02}}{1+\mathfrak{R}^{1.02}} \quad \text{and} \\
\mathfrak{R} &= \frac{r}{a_3}\,.
\end{align*}

The parameters in Eqs. \ref{eq:pbulge}, \ref{eq:pdisk} and \ref{eq:phalo} were taken from
\cite{allen} and they are listed in Table \ref{tab:params}. Although the model introduced by
\cite{allen} does not precisely  represent the current estimates of the mass distribution in the
Galaxy, this model has been widely used in studies of orbits of open clusters \citep{allen06,
bellini10} and in studies of moving groups in the solar neighbourhood \citep{antoja09, antoja}.
Moreover, \cite{jilkova12} did not find substantial differences in the orbit of an open cluster when
the axisymmetric component is described by a different more up-to-date  model. Therefore, we do not expect that the
modelling of the axisymmetric component of the Galaxy influences the results obtained in this study.

\subsection{Galactic bar}
\label{sect:bar}

We model the  bar of the Galaxy with a  three-dimensional Ferrers potential \citep{ferrers}, which is represented by the following density: 

\begin{equation}
  \rho_\mathrm{bar}= 
  \begin{cases}
    \rho_0 \left( 1-n^2 \right)^k & n < 1\\
    0 & n \geq 1
  \end{cases}\,.
  \label{eq:bar}
\end{equation}
The quantity $n$ determines the shape of the bar, which is given by the equation: $n^2= x_\mathrm{rot}^2/a^2 + y_\mathrm{rot}^2/b^2+ z_\mathrm{rot}^2/c^2$,  where the parameters $a$, $b$ and $c$ are the semi-major, semi-minor and vertical axes of the bar, respectively.  The term $\rho_0$ in Eq. \ref{eq:bar}  represents the central density of the bar and $k$ its concentration. Following \cite{merce2}, we chose $k=1$. 

The parameters that describe the bar such as its pattern speed, mass, orientation and axes are
currently under debate \citep[for a complete discussion see e.g.][]{martinezb14}. Hence, we used
values that are within the ranges reported in the literature. These values are listed in Table
\ref{tab:params}. 

\subsection{Spiral arms}
\label{sect:sp}

The spiral arms are usually represented as periodic perturbations of the axisymmetric potential. We use the prescription given by \cite{cox02}, which models such perturbations in the three-dimensional space. The potential of the spiral arms is given by the following expression:

\begin{equation}
\begin{split}
\Phi_\mathrm{sp}=& -4\pi GHA_\mathrm{sp}\exp{\left(-\frac{r_\mathrm{rot_1}}{R_\Sigma}
\right)}\sum\limits_n\left(\frac{C_n}{K_nD_n}\right)\times \\
&\cos({n\gamma}) \left[\mathrm{sech}\left( \frac{K_nz_\mathrm{rot_1}}{\beta_n}\right) \right]^{\beta_n}\,, 
\end{split}
\label{eq:sp}
\end{equation}
where $r_\mathrm{rot_1}$ is the distance of the star from the Galactic centre, measured in the frame co-rotating with the spirals arms. The value $H$ is the scale height, $A_\mathrm{sp}$ is the amplitude of the spiral arms and $R_\Sigma$ is
the scale length of the drop-off in density amplitude of the arms. We use $n=1$ term only, with $C_1=8/3\pi$ and the parameters $K_1, D_1$ and
$\beta_1$ given by:

\begin{align*}
K_1 &= \frac{m}{r_\mathrm{rot_1}\sin i},\\
\beta_1 &= K_1H (1+0.4K_1H), \\
D_1&= \frac{1+K_1H+0.3(K_1H)^2}{1+0.3K_1H}\,,
\end{align*}
where $m$ and $i$ correspond to the number of arms and pitch angle of the spiral structure
respectively. 

Finally, the term $\gamma$ in Eq. \ref{eq:sp} represents the shape of the spiral structure, which is
described by the expression:

\begin{equation*}
\gamma= m\left[ \varphi -\frac{\ln(r_\mathrm{rot1}/r_0)}{\tan i} \right]\,.
\end{equation*}
Here $r_0$ is a parameter which determines the scale length of the spiral arms. Following
\cite{jilkova12}, $r_0=5.6$~kpc.  

As for the bar, the parameters that describe the spiral structure of the Galaxy are rather uncertain
\citep[See e.g.][]{jilkova12, martinezb14}. Therefore, we chose the values that are consistent with
the current determination of the spiral structure. These values are listed in table \ref{tab:params}.

\subsection{Solar orbits}

\begin{figure}
  \centering
  \includegraphics[width=84mm]{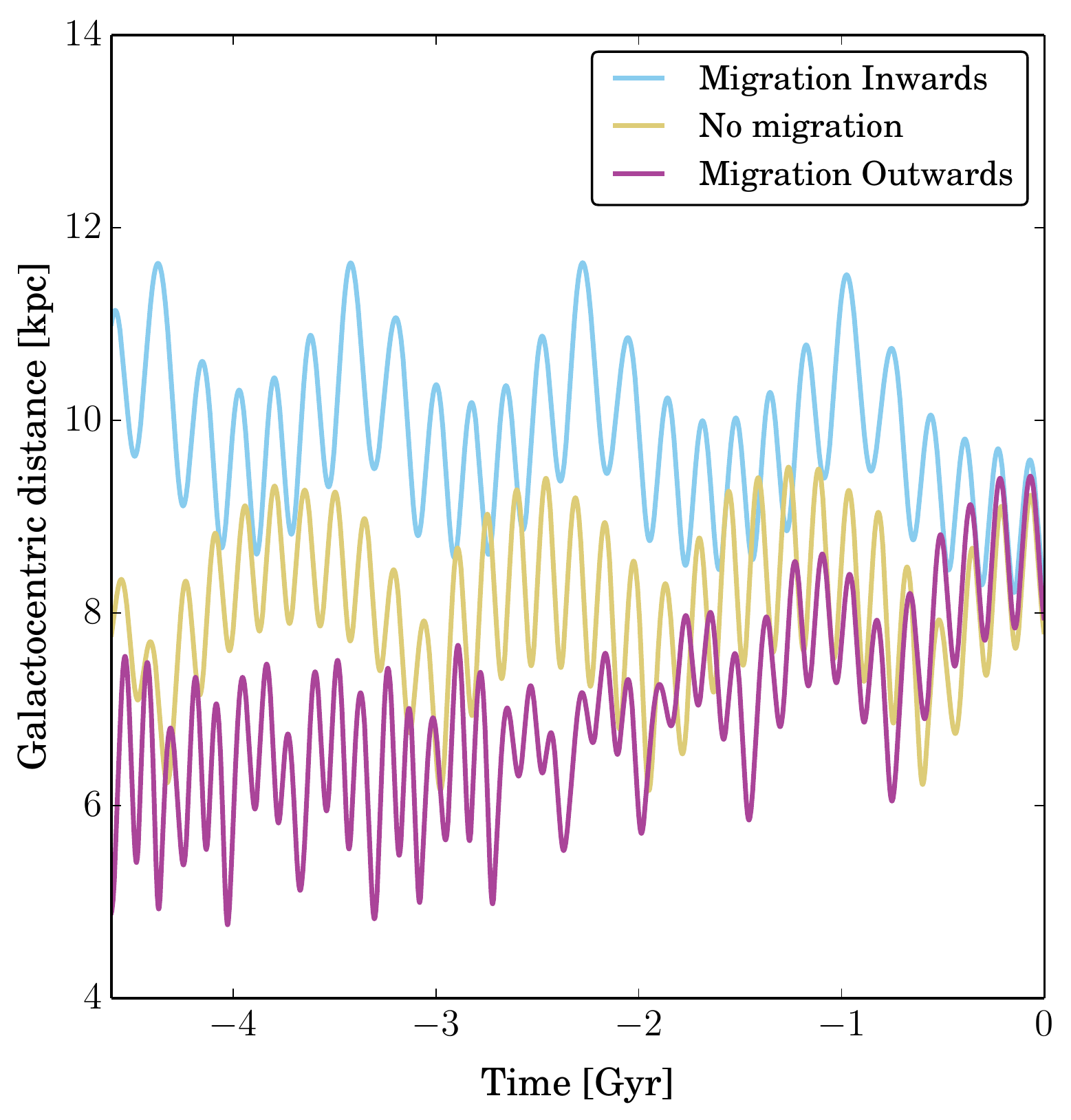}
  \caption{Possible trajectories of the Sun under the Galactic parameters listed in Table. \ref{tab:params}.  $t= 0$~Gyr represents the current time. \label{fig:orbits}}  
  \end{figure}

We calculate the  orbit of the Sun backwards in time using the analytical Galaxy model described
previously. In this calculation we account for the uncertainty in the present-day Galactocentric
phase-space coordinates of the Sun. We employ the same methodology as used by  \cite{martinezb14}
for this purpose\footnote{Unlike \cite{martinezb14} we use a three-dimensional model for the Galaxy
in this study; see Sect. \ref{sect:sp}}. Thus, we select a sample of $5000$ random positions and velocities from a normal
distribution centred at the current phase-space coordinates of the Sun.  The normal distribution is
then centred at $(r_\odot, v_\odot)$ with  standard deviations $(\sigma)$ corresponding to the
uncertainties in these coordinates.  In an inertial frame that is fixed at the center of the Galaxy, the present-day location of the Sun is (see Fig. \ref{fig:pos_sun}) :

\begin{align*}
r_\odot &= (-8.5, 0, 0.02)~\mathrm{kpc} \quad \text{and}\\
\sigma_r &= (0.5, 0, 0.005)~\mathrm{kpc},
\end{align*}
where the position of the Sun in the plane is given by: $R_\odot~=~8.5$~kpc. 

The present-day velocity of the Sun is: 

\begin{align*}
v_\odot &= (11.1, 12.4 +V_\mathrm{LSR}, 7.25)~\velocity \quad \text{and} \\
\sigma_v&= (1.2, 2.1, 0.6)~\velocity\,.
\end{align*}
where $v_\odot$ and $\sigma_v$ were taken from \cite{schonrich} and $V_\mathrm{LSR}$ corresponds to
the velocity in the Local Standard of Rest. According to the Milky Way model parameters listed in
Table \ref{tab:params}, $V_\mathrm{LSR}=226$~\velocity.

We integrate the orbit of the Sun backwards in time using each of the $5000$ positions and
velocities as initial phase-space coordinates.  The solar orbits were computed during $4.6$~Gyr by
using a  6th-order integrator called Rotating \bridge\ \citep[Pelupessy et al. in
prep.]{martinezb14}.  This integrator is implemented in the \amuse\ framework \citep{portegies13, pelupessy13}.


At the end of the calculation, we obtain a collection of solar orbits, from which we chose three.
These orbits are shown in Fig.~\ref{fig:orbits} and they represent different orbital histories of
the Sun through the Galaxy. The blue orbit for instance shows that the Sun might have been born at
$\sim 11$~kpc from the Galactic centre, suggesting migration from outer regions of the Galactic disk
to $R_\odot$. \cite{martinezb14} argued that such a migration could only have happened if the Sun
was influenced by the overlapping of the co-rotation resonance of the spiral arms with the Outer
Lindblad resonance of the bar. On the other hand, the violet orbit shows an example where the Sun migrated
from inner parts of the disk to $R_\odot$, in accordance with \cite{wielen96} and \cite{minchev13}.
The yellow orbit represents the case where the Sun does not migrate on average.

The stellar encounter rate experienced by the Sun during the last $4.6$~Gyr depends on the solar orbit, due to differences in the stellar density and in the local stellar velocity dispersion. Therefore we compute the number of stellar encounters in each of the orbits shown in Fig. \ref{fig:orbits}.  The methodology is described in Sect. \ref{sect:nenc}.

\section{Galactic stellar encounters}
\label{sect:nenc}

The frequency of stellar passages along the orbit of the Sun, $f$, can be estimated by the following equation \citep[][]{garcia01}:

\begin{equation}
f= \sum_i f_i= \pi D^2\sum_i n_iv_i.
\label{eq:frec}
\end{equation}

\noindent
The index $i$ denotes different stellar types according to the classification given in \citet[][Table 8]{garcia01}. The term $D$ corresponds to the maximum pericentric distance from the Sun where a stellar encounter is considered. We set $D= 4\times10^5$~AU, because we do not expect farther encounters to substantially perturb the Solar system \citep[see e.g.][]{rickman08, 2014MNRAS.442.3653F}. The quantity $n_i$ in Eq. \ref{eq:frec} corresponds to the number density of each stellar type,  along the orbit of the Sun. The term $v_i$ is the velocity of the encounter which is described by the expression:

\begin{equation}
v_i= \left[ v_{\odot i}^2 + \nu_i^2 \right]^{1/2}.
\label{eq:venc}
\end{equation}

Here $v_{\odot i}$ corresponds to  the Sun's peculiar velocity relative to the star belonging to the $i$-th category (we assume that $v_{\odot i}$ is constant everywhere in the Galaxy).  The term $\nu_i$ is the velocity dispersion of the given stellar type, along the orbit of the Sun. 

We obtain $n_i$ and $\nu_i$ by using a similar procedure briefly described in  \cite{kaib11}. 
In Sects. \ref{sect:density} and \ref{sect:dispersion} we explain this methodology in more detail.

\subsection{Estimation of $n_i$}
\label{sect:density}

We obtain the number density of a given stellar type along the orbit of the Sun, $n_{i}$ by scaling up or down the number density of that stellar type at the current solar position, $n_{i\odot}$ . The number density $n_i$ is therefore  given by the following expression:

\begin{equation}
n_i= \beta n_{i\odot}.
\label{eq:ni}
\end{equation}  

We take the values of $n_{i\odot} $ from \citep[][Table8]{garcia01}. The quantity $\beta$ is a scaling factor that depends on the location of the Sun in the Galaxy.  We compute $\beta$  by assuming that the number densities have the same spatial distribution through the Galaxy \citep[see e.g.][]{2014MNRAS.442.3653F}. The scaling factor is therefore equals to:

\begin{equation}
\beta= \frac{\rho}{\rho_\odot},
\end{equation}
where $\rho$ is the local stellar mass density along the Sun's orbit,  and $\rho_\odot$ is the local stellar density at the current Sun's position.  We compute $\rho$ and $\rho_\odot$  through the Poisson's equation using the Galaxy potential described in Sect.\ \ref{sect:gal_model}.  In the calculation of the local stellar mass density, we do not include the dark matter halo potential.

\subsection{Estimation of $\nu_i$}
\label{sect:dispersion}

We obtain the velocity dispersion of a given stellar type along the Sun's orbit, $\nu_i$ by scaling up or down the velocity dispersion of that stellar type at the current position of the Sun, $\nu_{i\odot}$. The velocity dispersion $\nu_i$ is described by the following expression:

\begin{equation}
\nu_i= \alpha \nu_{i\odot}.
\label{eq:sigmai}
\end{equation}

The values of  $\nu_{i\odot}$ are taken from \citet[][Table 8]{garcia01}. The scaling factor $\alpha$ depends on the location of the Sun in the Galaxy and it is given by:
\begin{equation}
\alpha= \frac{\nu}{\nu_\odot},
\end{equation}
where $\nu$ is the total velocity dispersion at a given location along the Sun's orbit and $\nu_\odot$ is the total velocity dispersion at the current position of the Sun. $\nu_\odot$ is the weighted average of the velocity dispersions per stellar type (the weights being equal to $n_{i\odot}$).

\begin{figure}
  \centering
  \includegraphics[width=84mm]{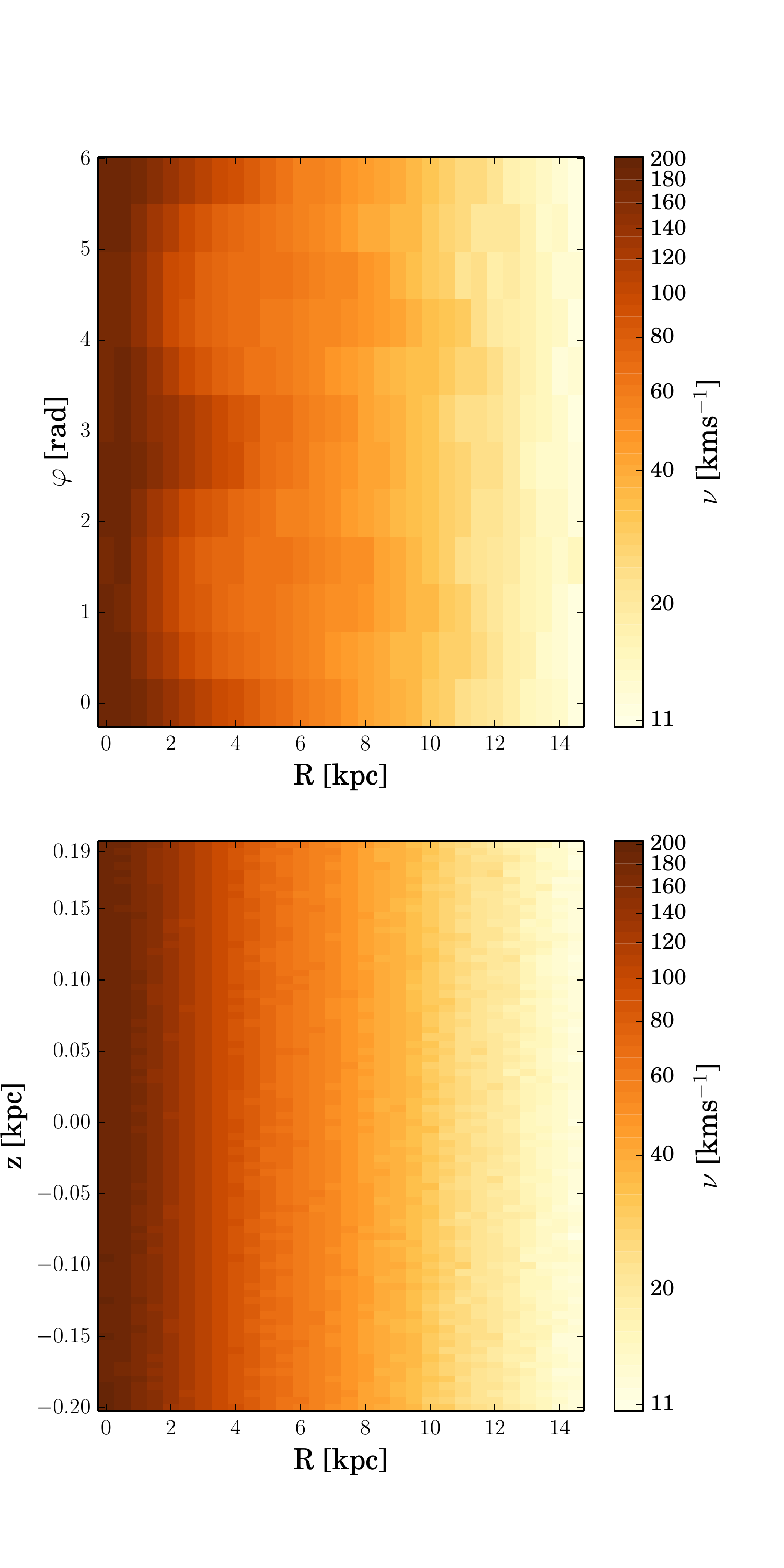}
  \caption{ {\it Top:} Stellar velocity dispersion of the Milky Way as a function of Galactocentric
  radius and azimuth where $0 \leq z \leq5$~pc. {\it Bottom:} Stellar velocity dispersion as a
  function of Galactocentric radius and vertical distance where  $0 \leq \theta \leq \pi/6$~rad.
 \label{fig:sigma_MW}}  
  \end{figure}

We obtain $\nu$ by using the largest N-body simulation of the Milky Way, which employs a total number of  $51$ billion particles \citep[][]{bedorf}.  We did not use the Galactic model described in Sect. \ref{sect:gal_model} given the complexity in the estimate of $\nu$ from an analytical Galaxy model. Although the computation of $\nu$ by means of a different Galaxy model is not consistent,  we note that the simulations of \cite{bedorf} have successfully reproduced the stellar velocity distribution within $500$~pc from the Sun (see e.g. Fig. 3 in their paper).  

We compute $\nu$ by using the snapshot of the simulation of \cite{bedorf} corresponding to $5.6$~Gyr of evolution. We chose this snapshot because it corresponds well to the current picture of the Milky Way.  In this snapshot, we discretize the space in bins of $(\Delta_\mathrm{R}, \Delta_\varphi, \Delta_z)=(0.3~\text{kpc}, 0.26~\text{rad},  5~ \text{pc})$ respectively.  This choice ensures a robust estimate of $\nu$ because of the number of particles located at each bin. The region in the Galaxy where we determine $\nu$ is: $0\leq R\leq 15$~kpc; $0\leq\varphi\leq 2\pi$~rad and  $-200\leq z\leq 200$~pc. 

The velocity dispersion in the $j$-th bin is given by the following expression:
\begin{equation}
\nu^2_j=  \frac{1}{N_j-1}\sum\limits_{k=1}^{N_j} \biggl [(v_{\mathrm{R}_{kj}}-\bar{v}_{\mathrm{R}j})^2 + (v_{\varphi_{kj}}-\bar{v}_{\varphi j} )^2 +(v_{z_{kj}}-\bar{v}_{zj})^2 \biggr] ,
\end{equation}
where $v_{\mathrm{R}_{kj}}, v_{\varphi_{kj}}$ and $v_{z_{kj}}$ are the radial, tangential and vertical velocities of the $k$-th star in the $j$-th bin that contains $N_j$ stars. $\bar{v}_{{\mathrm{R}j}}, \bar{v}_{\varphi j}$ and $\bar{v}_{zj}$ are the mean values of the former velocities respectively.

In Fig. \ref{fig:sigma_MW} we show $\nu$ as a function of the radius and azimuth (top panel) and as
a function of the radius and vertical distance (bottom panel).  As is expected, the velocity
dispersion decreases with radius, due to a reduction of the stellar density in the outer regions of
the Galaxy. At the solar position, we  observe that $\nu\simeq~40$~km~s$^{-1}$, which is in
agreement with measurements  of the local velocity dispersion \citep{nordstrom04,holmberg}.

The velocity dispersion varies periodically with azimuth, being higher in the inner disk (e.g. top panel Fig. \ref{fig:sigma_MW}). This variation is a signature of the presence of the bar which extends up to $\sim 4$~kpc from the Galactic centre.  The variation of $\nu$ with azimuth is smaller in outer regions of the disk and it is due to the presence of spiral arms.  From Fig. \ref{fig:sigma_MW} we also observe that the variation of the velocity dispersion with the vertical distance $z$ is low compared to the change with radius or  azimuth.

\subsection{Total number of encounters along the Sun's orbit} 
 
Once $n_i$ and $v_i$ are computed,  we can use Eq. \ref{eq:frec} to obtain the frequency of stellar encounters  experienced by the Sun along its orbit, $f$. Given that $f$ is a function of time (note that $n_i$ and $v_i$ change along the orbit), the total number of stellar encounters experienced by the Sun along its orbit is:

\begin{equation}
n_\mathrm{enc}= \int_{t=0}^{4.5~\text{Gyr}} f(t) \mathrm{d}t.
\label{Eq:nenc}
\end{equation}

For the solar orbit where the migration is inwards, $n_\mathrm{enc}~\backsimeq~9.3~\times~10^4$. For the solar orbit with migration outwards, $n_\mathrm{enc}~\backsimeq~28.2\times10^4$. For the orbit with no net migration,  $n_\mathrm{enc}~\backsimeq~17.5\times10^4$. We note that this last value is similar to that obtained by \cite{rickman08, 2014MNRAS.442.3653F}, who assumed a non-migrating orbit for the Sun (they found $n_\mathrm{enc}=197906$).

For each of the solar orbits shown in Fig. \ref{fig:orbits},  we generate a sample of $n_\mathrm{enc}$ random stellar encounters. The properties of these encounters -- time of occurrence (\tenc); mass (\menc); pericenter distance (\renc) and velocity (\venc) -- are calculated as explained bellow.

\begin{table}
  \caption{Mass ranges (\menc) corresponding to the magnitude intervals ($M_\mathrm{V}$) of \citet{garcia01}.
  The mass intervals for types B0--M5 are based on \citet{2013ApJS..208....9P}, \citet{2012ApJ...746..154P}, and \citet{2016_mamajek}, on \citet{2007MNRAS.375.1315K} for white dwarfs (WD), and on \citet{1973asqu.book.....A} for the giants.}
  \label{tab:mass_enc}
  \begin{tabular}{l l@{\hspace{2mm}}l l@{\hspace{2mm}}l}
    \hline \\[-1.7ex]
    Stellar type        & \multicolumn{2}{l}{$M_\mathrm{V}$ [mag]}  & \multicolumn{2}{l}{$M\enc$ [M$_{\sun}$]}  \\[0.5ex] \hline \\[-1.7ex]
    B0      & $-5.7$ & $-0.2$   & $60$   & $3.4$    \\
    A0      & $-0.2$ & $1.3$    & $3.4$  & $2.15$   \\
    A5      & $1.3$  & $2.4$    & $2.15$ & $1.67$   \\
    F0      & $2.4$  & $3.6$    & $1.67$ & $1.25$   \\
    F5      & $3.6$  & $4.0$    & $1.25$ & $1.18$   \\
    G0      & $4.0$  & $4.7$    & $1.18$ & $1.02$   \\
    G5      & $4.7$  & $5.5$    & $1.02$ & $0.9 $   \\
    K0      & $5.5$  & $6.4$    & $0.9 $ & $0.78$   \\
    K5      & $6.4$  & $8.1$    & $0.78$ & $0.64$   \\
    M0      & $8.1$  & $9.9$    & $0.64$ & $0.51$   \\
    M5      & $9.9$  & $18.0$   & $0.51$ & $0.082$  \\
    WD$^{a}$& ---    & ---      & \multicolumn{2}{l}{$\mu=0.59$, $\sigma=0.07$}\\
    Giants  & ---    & ---      & $2.5$  & $6.3$ \\[0.5ex]
    \hline
  \end{tabular} \\[0.5ex]
  $^{a}$ In the case of white dwarfs, the listed numbers $\mu$ and $\sigma$ correspond to the mean and standard deviation of the Gaussian distribution, respectively.
\end{table}

We randomly draw $t_\mathrm{enc}$ with a probability that is proportional to the encounter frequency. Once we determine \tenc, we proceed to the computation of 
the mass of the encounters, \menc. This quantity is sampled by using the data listed in 
\citet[][Table~8]{garcia01} which comprises the properties of different stellar types defined by intervals in visible magnitude. First we determine the $i$-th stellar type  of each encounter according to the number density $n_i$ \footnote{$n_i$ depends on \tenc, since $n_i$ is the stellar density measured along the Sun's orbit (Eq. \ref{eq:ni}).}.
The mass is determined for each encounter as follows.
For stellar types A0--M5, we define mass ranges corresponding to the magnitude intervals based on \citet{2013ApJS..208....9P}, \citet{2012ApJ...746..154P}, and \citet{2016_mamajek}\footnote{We used data compiled by \citet{2016_mamajek} and publicly available at the web page \url{http://www.pas.rochester.edu/~emamajek/EEM_dwarf_UBVIJHK_colors_Teff.txt}}. We list the magnitude and mass in Table~\ref{tab:mass_enc}.
We pick the  individual masses from the mass range of the corresponding stellar type and with distribution given by \citet{kroupa}
\begin{equation}
  \frac{\mathrm{d}N}{\mathrm{d}M} \propto
  \begin{cases}
    M^{-1.3}, & 0.08 < M \leq 0.5\,\mathrm{M}_{\sun}, \\
    M^{-2.3}, & 0.5  < M < 60\,\mathrm{M}_{\sun},
  \end{cases}
\end{equation}
where we chose the maximum mass of 60\,$\mathrm{M}_{\sun}$ (the results are not sensitive to the upper limit since the frequency of B0 stars is very low).

For white dwarfs, we assume a Gaussian distribution with the mean of 0.59\,$\mathrm{M}_{\sun}$ and standard deviation of 0.07\,$\mathrm{M}_{\sun}$. We derive these values as means of the four distributions in Table~1 of \citet{2007MNRAS.375.1315K}, weighted by fraction of stars in their sub-samples.
Finally, we take the limiting masses for giant stars from \citet[][as the masses of G0 and M0 giants]{1973asqu.book.....A} and we assume uniform distribution between 2.5 and 6.3\,$\mathrm{M}_{\sun}$.

We generate the distribution of encounter velocities, \venc\ by adopting the same methodology as  \cite{2014MNRAS.442.3653F}. The procedure is as follows. The magnitude of the encounter velocity in the heliocentric reference frame is:

\begin{equation}
v_\mathrm{enc}= \left[ v_{\odot i}^2 + V_i^2 -2v_\odot V_i\cos\delta \right]^{1/2}.
\label{eq:Venc}
\end{equation}
Here, $v_{\odot i} $ is the solar apex velocity relative to the star of $i$-th ca\-tegory (note that the stellar category was previously chosen from $n_i$). The term $V_i$ is the velocity of the stellar encounter in the Local Standard of Rest (LSR). $\delta$ is the angle between $v_{\odot i}$ and $V_i$ in the LSR.

The velocity of the stellar encounter in the LSR is given by the following equation:

\begin{equation}
V_i= \nu_{i}\left[\frac{1}{3}\left(\eta _u^2+ \eta_v^2 +\eta_w^2 \right) \right]^{1/2},
\label{eq:Vi}
\end{equation}
where the quantities $\eta _u$, $\eta_v$, $\eta_w$ are random variables that follow a Gaussian distribution with zero mean and unit variance. We obtain the distribution of $v_\mathrm{enc}$ in the following way:  i) we randomly generate $\cos\delta$ from a uniform distribution in the range [$-1$, $1$].  ii) Adopting $v_{\odot i}$ from \cite[][Table8]{garcia01} and  computing $\nu_i$ from Eq. \ref{eq:sigmai}, we calculate $V_i$ from Eq. \ref{eq:Vi} and  $v_\mathrm{enc}$ using Eq. \ref{eq:Venc}. iii) Since we have to account for the fact that the contribution to the encounter flux is proportional to $v_\mathrm{enc}$, we define a large velocity, $V_\mathrm{enc}= v_{\odot i}+ 3\nu_{i}$. iv) According to the stellar category, we randomly draw a velocity $v_\mathrm{rand}$ from an uniform distribution over $[0, V_\mathrm{enc}]$. If $v_\mathrm{rand}<v_\mathrm{enc}$, we accept $v_\mathrm{enc}$ and the generated values of $\cos\delta$, $V_i$. Otherwise, we reject it and repeat the process until $v_\mathrm{rand}<v_\mathrm{enc}$.

Finally, we sample the distances of the stellar encounter, \renc\ from a  distribution function proportional to $r_\mathrm{enc}$ with an upper limit of $4\times 10^5$~AU, in the same fashion as \cite{2014MNRAS.442.3653F}.

\begin{figure}
  \centering
  \includegraphics[width=84mm]{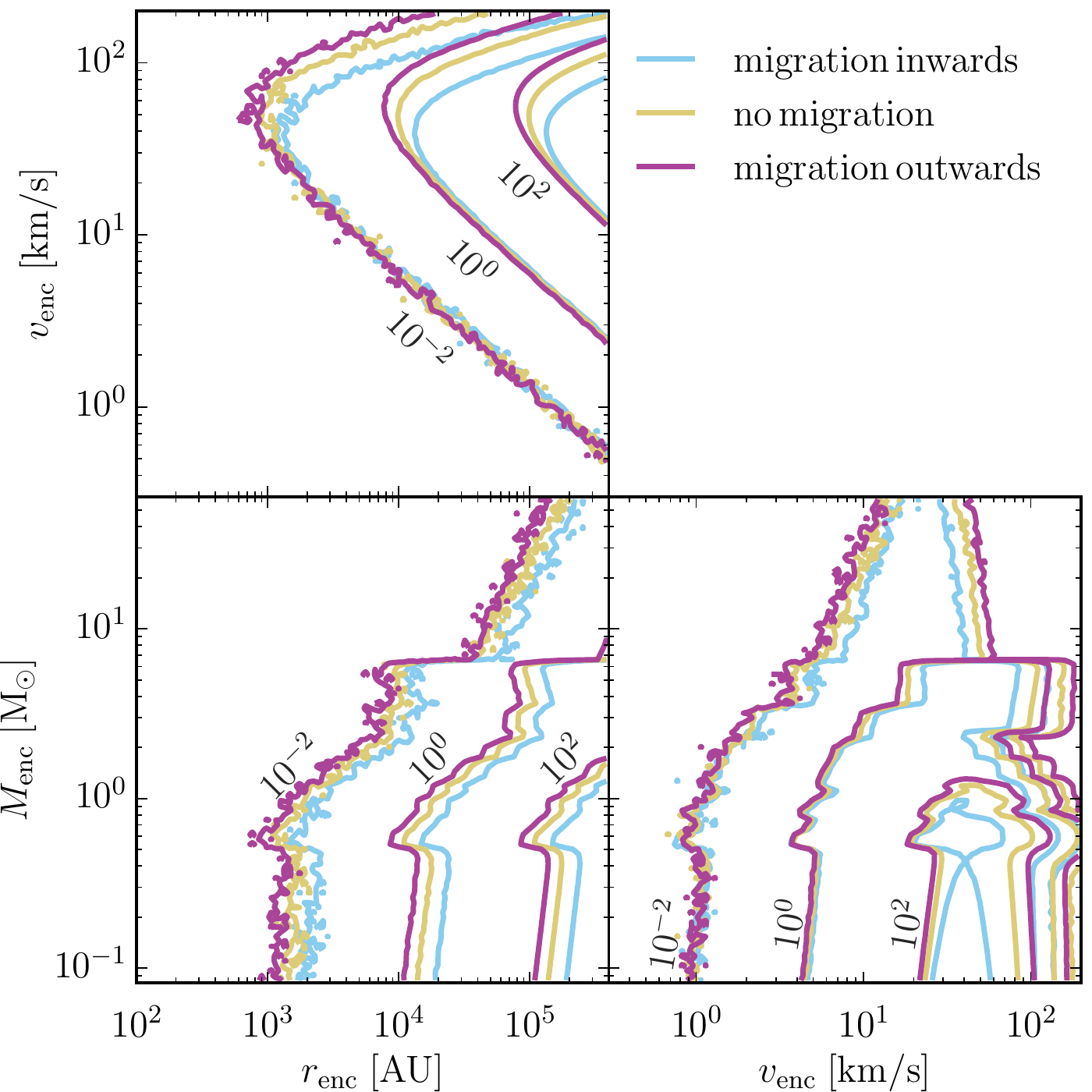}
  \caption{
 Number of encounters as a function of the mass $M\enc$, velocity $v\enc$, and pericenter $r\enc$ of the encountering star along the thee studied orbits.
  The number of encounters, $n\enc$, is averaged over the number of generated sets ($1000$, see text).
  In each subplot, three contours ($n\enc=10^{-2}, 10^{0}, 10^{2}$ per bin) of different two-dimensional distributions are shown.
  The axes are logarithmic and $n\enc$ is not normalized by the size of the bin. Hence, the plot serves for the comparison between the three different orbits.
   } \label{fig:encounters_mrv}
\end{figure}

For each of the three studied orbits, we calculated and combined 1000 different sets of encounters (realized by different random seeds) following the method described above.
In Fig.~\ref{fig:encounters_mrv} we show the distributions of the encounters averaged over the total number of sets (1000) in two-dimensional projections of the space of M\enc, v\enc\ and r\enc. Note that the distributions do not differ dramatically with migration.
As expected from the assumed distributions, most of the encounters are with low-mass stars ($M\enc<1\,\mathrm{M}_{\sun}$) and velocities of $\sim$20--100\,km\,s$^{-1}$.

From the large set of stellar encounters obtained, we can look for those that produce the strongest perturbation in the  outer regions of the Solar system. 
These stellar encounters will set the outer edge of the parking zone. \cite{portegies15}, used the encounter with Scholz's star to determine the location of the outer edge of the Solar system's parking zone,  they found that the effect of this particular encounter has hardly perturbed the Oort cloud down to a distance of $10^5$~AU. If the Sun experienced  stronger stellar encounters, the perturbations might become important at smaller semi-major axes, shifting inwards the outer edge of the parking zone. In the next section we determine the strongest stellar encounters  experienced by the Sun and we make a new estimate of the location of the outer edge of the Solar system's parking zone. \\

\section{The outer limit of the parking zone}
\label{sect:stability}

We estimate the outer limit of the parking zone using the impulse approximation \citep{rickman76}.
The impulse approximation assumes that the velocity vector of the perturbing star, $\mathbfit{v}\enc$, and the position vector of the perturbed body orbiting the Sun are constant during the encounter.
This corresponds to the assumption that the timescale of the encounter is much longer than the orbital period of the perturbed body.
Following \citet{portegies15}, we further assume that the point of the closest approach of the star lies on the line joining the Sun and the perturbed body (that is the velocity of the perturbing star $\mathbfit{v}\enc$ is perpendicular to the position vector of the perturbed body), which is the geometry resulting in the maximal perturbation.
Finally we assume that the perturbed body is at the aphelion of its orbit where it is moving the slowest.

The impulse gained by a perturbed body moving on an orbit with semi-major axis $a$ and eccentricity $e$ then is
\begin{equation}\label{eq:impulse}
\Delta I=\frac{2\mathrm{G}M\enc}{v\enc r\enc}\frac{a(1+e)}{r\enc-a(1+e)}.
\end{equation}
Note that in the case of a distant encounter, when $r\enc\gg a(1+e)$, the impulse given in Eq.~\ref{eq:impulse} at given distance from the Sun is proportional to $M\enc/(v\enc r\enc^2)$.
\citet{2015MNRAS.454.3267F} used this expression as a proxy for the strength of the encounters (as measured by the number of injected long-period comets).
We define the outer limit of the parking zone where the perturbation corresponds to the body's velocity at aphelion \citep{portegies15}, that is 
\begin{equation}\label{eq:v_apo}
\Delta I=\sqrt{\frac{\mathrm{G}M_{\sun}}{a}\frac{1-e}{1+e}},
\end{equation}
where the mass of the Sun is $M_{\sun}=1\,\mathrm{M}_{\sun}$.
From Eqs.~\ref{eq:impulse} and~\ref{eq:v_apo}, we can find the semi-major axis of the outer limit of the parking zone as a function of eccentricity, $a_{\mathrm{PZ}}(e)$.

\begin{figure}
  \centering
  \includegraphics[width=84mm]{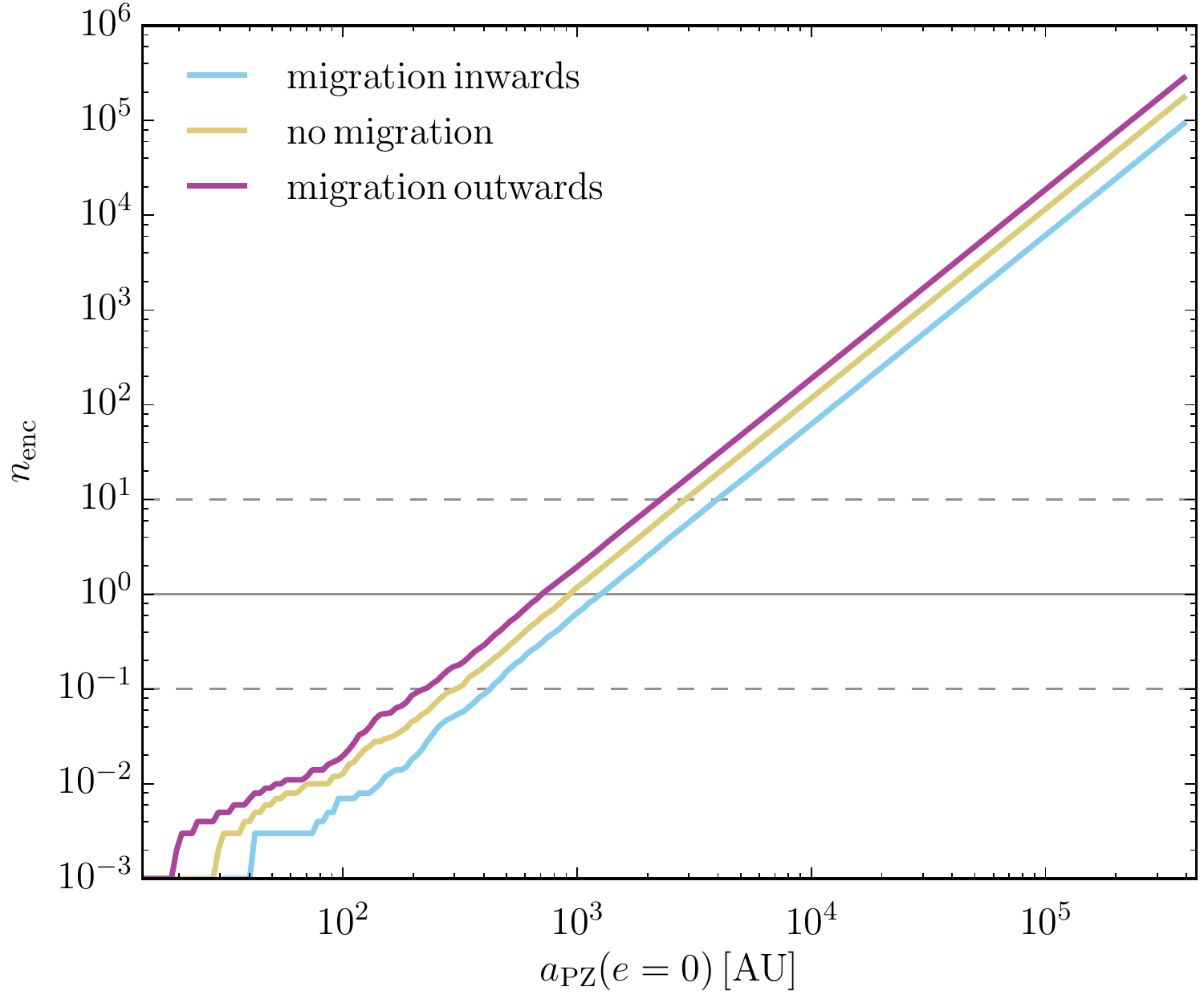}
  \caption{
  Cumulative distributions of the outer limit of the parking zone for circular orbits $a_{\mathrm{PZ}}(e=0)$.
  The three orbits with different migration are shown.
  The distributions are derived using 1000 different encounter sets (corresponding to different random seeds) and the number of encounters, $n\enc$, is averaged over these number of sets.
  The maximal value of $a_{\mathrm{PZ}}(e=0)$ is given by the upper limit of the encounter pericenter of $D=4\times10^5$\,AU (Sect.~\ref{sect:nenc}).
  The maximal value of $n\enc$ corresponds to the total number of encounters along the orbits.
  The horizontal lines indicate $n\enc=0.1$, 1, and 10.
  Both horizontal and vertical axes are logarithmic.
  } \label{fig:pz_e0_cumulative}
\end{figure}

In Fig.~\ref{fig:pz_e0_cumulative}, we compare cumulative distributions of the outer limit of the parking zone of a circular orbit, or $a_{\mathrm{PZ}}(e=0)$, for encounters along each of the studied orbits.
To obtain the distributions, we generated 1000 different sets of encounters for each solar orbit and calculated their $a_{\mathrm{PZ}}(e=0)$.
The distributions in Fig.~\ref{fig:pz_e0_cumulative} are averaged over the number of encounter sets and $n\enc$ is the number of encounters per orbit.

\begin{figure}
  \centering
  \includegraphics[width=84mm]{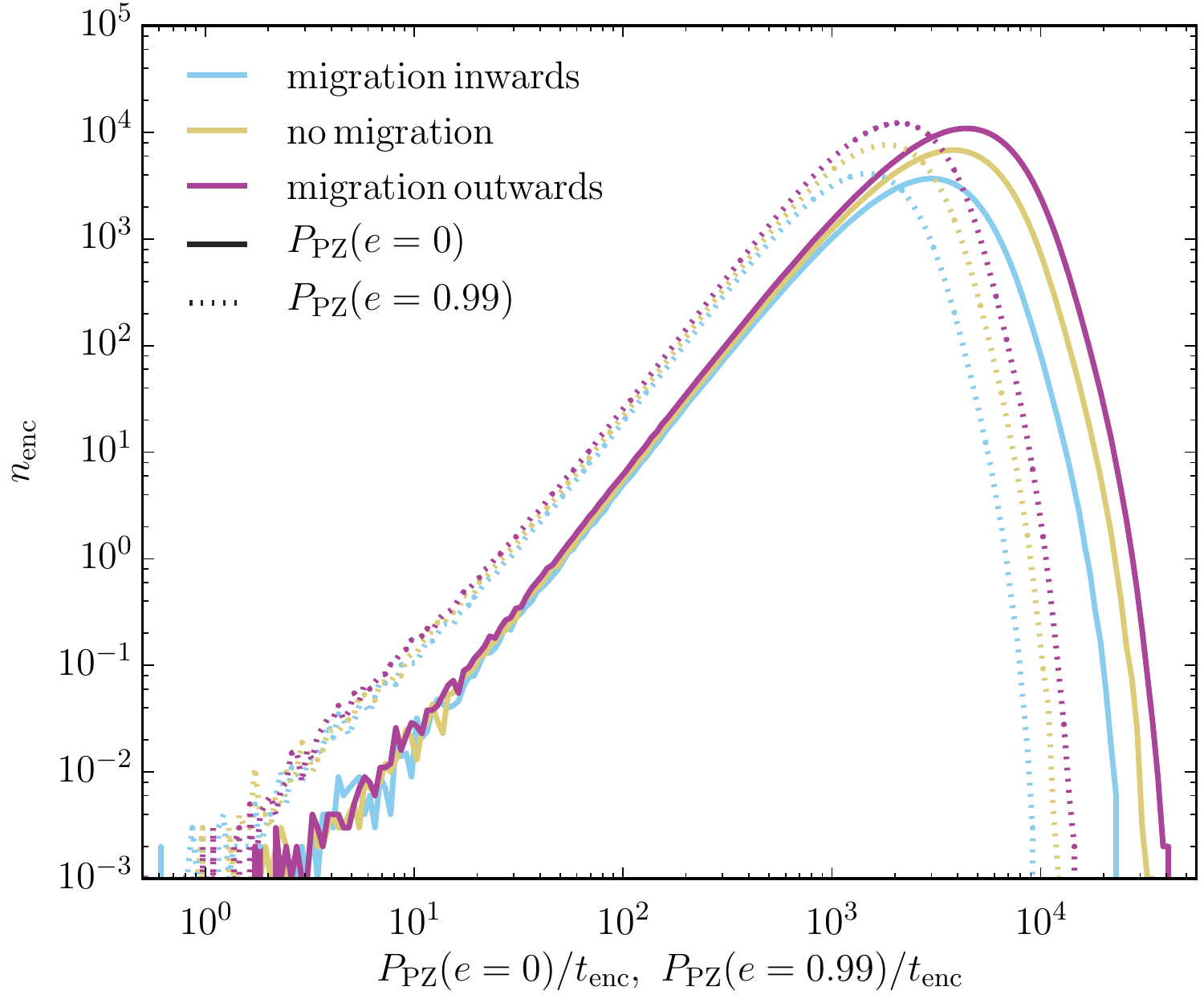}
  \caption{
  Distributions of the ratio of the period of a circular orbit of semi-major axis $a_{\mathrm{PZ}}$ for circular and eccentric ($e=0.99$) orbits, $P_{\mathrm{PZ}}(e=0)$ and $P_{\mathrm{PZ}}(e=0.99)$, and the timescale $t\enc$ of the corresponding encounter.
  The three different orbits are shown by different colors.
  Full and dotted lines correspond to circular and eccentric orbits respectively.
  The distributions are derived for the same combined encounter sets as in Fig.~\ref{fig:pz_e0_cumulative}. 
Note that  both horizontal and vertical axes are logarithmic.
  } \label{fig:pPZ_tenc_ratio}
\end{figure}

The impulse approximation is based on the assumption that the timescale of the encounter is much shorter than the orbital period of the perturbed body.
To verify the validity of this assumption, we compare the period $P_{\mathrm{PZ}}(e=0)$ of the circular orbits with semi-major axes of $a_{\mathrm{PZ}}(e=0)$ with the timescales of the encounters taken as $t\enc=r\enc/v\enc$.
The distributions of the ratio $P_{\mathrm{PZ}}(e=0)/t\enc$ are shown in Fig.~\ref{fig:pPZ_tenc_ratio} by full lines. We found that
for the vast majority of the encounters (more than 99\%), the ratio is higher than 100 and the assumption is well fulfilled.
There is only a very small number of encounters (less than few dozen in the combined encounter sets, translating into $n\enc\sim2\times10^{-2}$ per orbit in Fig.~\ref{fig:pz_e0_cumulative}), typically close ($r\enc\lesssim100$\,AU) and slow ($v\enc\lesssim10$\,km\,s$^{-1}$), for which $P_{\mathrm{PZ}}(e=0)/t\enc\lesssim10$.
Given that there is less than 1\% of encounters with $P_{\mathrm{PZ}}(e=0)/t\enc<50$, we neglect the inaccuracy of the impulse approximation for these cases.

The outer limit of the parking zone for higher eccentricities reaches smaller semi-major axes ($a_{\mathrm{PZ}}$ decreases with eccentricity) than for the circular orbits.
The dotted lines in Fig.~\ref{fig:pPZ_tenc_ratio} show the distributions of the ratio $P_{\mathrm{PZ}}(e=0.99)/t\enc$ for the three diffe\-rent orbits.
Note that the distributions are shifted to lower values compared to $P_{\mathrm{PZ}}(e=0)/t\enc$ (full lines). For an eccentricity of $e=0.99$, we found that there is about 17\% of encounters with $P_{\mathrm{PZ}}(e=0.99)/t\enc<50$.
The overall fraction of encounters with $P_{\mathrm{PZ}}(e=0.99)/t\enc<50$ however, is still relatively small, typically of the order of $10^{-4}$ out of the total number of encounters.
More accurate approximations \citep{1994CeMDA..58..139D,2005EM&P...97..411R} can be use to remedy this inaccuracy, but here we stick to the impulse approximation.

\begin{table}
  \caption{Stellar encounters used for the parking zone's outer limit in Fig.~\ref{fig:pz}.}
  \label{tab:pz_parameters}
  \begin{tabular}{l l l l}
    \hline \\[-1.7ex]
    Orbit               & $M\enc$ [M$_{\sun}$]  & $r\enc$ [AU]  & $v\enc$ [km\,s$^{-1}$] 
                        \\[0.5ex] \hline \\[-1.7ex]
    Migration inwards   & $0.11$            & $1285$        & $38.5$    \\
    No migration        & $0.13$            & $948$         & $30.4$    \\
    Migration outwards  & $0.43$            & $721$         & $21.4$    \\[0.5ex]
    \hline
\end{tabular}
\end{table} 

\begin{figure}
  \centering
  \includegraphics[width=84mm]{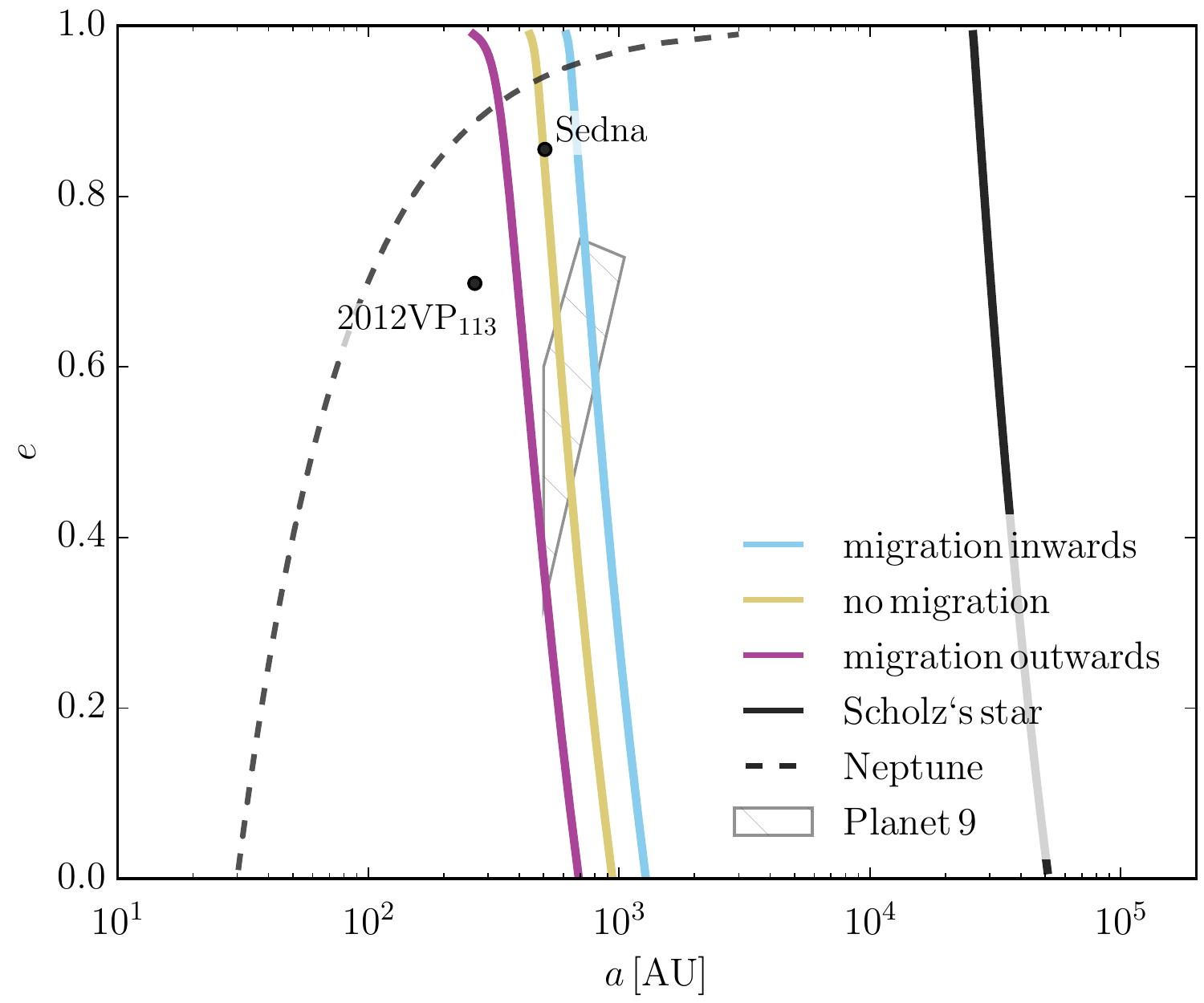}
  \caption{
  Solar system's parking zone in the plane of eccentricity $e$ and semi-major axis $a$.
  Its inner limit is defined by the perturbations by Neptune and indicated by dashed black line.
  The blue, yellow and purple lines show its outer limit for solar orbits with different migration (see Fig.~\ref{fig:orbits} and Table~\ref{tab:pz_parameters}).
  The black line is the estimate of the outer limit using Scholz's star \citep{portegies15}.
  The two bullet points indicate the orbital elements of the inner Oort cloud bodies Sedna \citep{brown04} and 2012VP$_{113}$ \citep{trujillo}.
  The hashed triangle shows the orbital elements constrains for Planet\,9 \citep{2016ApJ...824L..23B}; see Sect.~\ref{sect:discussion3} for discussion.
} \label{fig:pz}
\end{figure}

We determine the actual outer edge of the Solar system's parking zone such that the number of encounters along the orbit resulting in smaller $a_{\mathrm{PZ}}(e=0)$ is $n\enc=1$.
These values are marked by the horizontal full gray line in Fig.~\ref{fig:pz_e0_cumulative}.
We obtain $a_{\mathrm{PZ}}~(e~=~0)~\approx1280$, 940 and 690\,AU for the orbit with inwards migration, no migration, and outwards migration, respectively.
In Fig.~\ref{fig:pz} we show the resulting Solar system's parking zone.
We list the parameters of the encounters used to calculate the outer edge of the parking zone in Table~\ref{tab:pz_parameters}.
These encounters were determined from the cumulative distribution of the number of encounters (Fig.~\ref{fig:pz_e0_cumulative}) where we picked the encounters with the smallest $a_{\mathrm{PZ}}(e=0)$ of the first bin with $n\enc>1$.
Note that these are an example encounters and different combinations of parameters result in the same $a_{\mathrm{PZ}}(e=0)$ and parking zone's outer limit in the plane $e\times a$.
For fixed semi-major axis $a$ and eccentricity $e$, the parameters of the encounters resulting in the same change of impulse are bound as $M\enc/[v\enc r\enc (r\enc-a)]=\mathrm{const.}$ (Eq.~\ref{eq:impulse}).
We use the following parameters to draw the outer edge of the Solar system's parking zone:  $a=a_{\mathrm{PZ}}(e=0)$; $e=0$; $M\enc/[v\enc r\enc (r\enc-a_{\mathrm{PZ}})]=7.9$, $6.7$, and $2.2\times10^{-5}\mathrm{M}_{\sun}\,\mathrm{AU}^{-2}\,\mathrm{km}^{-1}\,\mathrm{s}$. The encounters with $M\enc$, $v\enc$ and  $r\enc$ that give these values will result in the same outer limit of the parking zone.

The solid black line in Fig.~\ref{fig:pz} corresponds to the original estimate made by \cite{portegies15}  using the Scholz's star.  
The outer edge of the parking zone given by the encounters derived here is located in the region corresponding to the inner Oort cloud, where objects like Sedna \citep{brown04} and $2012$VP$_{113}$ \citep{trujillo} reside.
The orbit migrating outwards from the denser inner regions of the Galaxy (violet line in Figs.~\ref{fig:orbits} and \ref{fig:pz}), results in the smallest parking zone.
The orbit migrating inwards from the less dense outer regions (blue line in Figs.~\ref{fig:orbits} and \ref{fig:pz}) results in a parking zone that would not perturb the objects on Sedna-like orbits.
This picture is consistent with \cite{kaib11}, who concluded that the inner edge of the classical Oort cloud strongly depends on the orbit of the Sun, being smaller for the orbits that moved closer to the Galactic center.

\section{discussion}
\label{sect:discussion3}

\subsection{Effect of stellar encounters on the hypothetical Planet 9}

In order to explain some of the observed characteristics of distant KBOs or inner Oort cloud bodies, there has been ongoing discussion on an undiscovered planet in the outer Solar system (for example \citealp{1985Natur.313...36W}, \citealp{1986Icar...65...37M}, \citealp{1999MNRAS.309...31M}, \citealp{2002MNRAS.335..641H}, \citealp{2004Icar..171..516M}, \citealp{2006Icar..184..589G}, \citealp{2008AJ....135.1161L}, \citealp{2015Icar..258...37G} and others).
The most recent prediction in this context was made by \citet{BB16} who showed that the presence of a distant planet\,--\,so-called Planet~9 (hereafter P9), can explain the observed orbital alignment of some KBOs and inner Oort cloud objects. 

\citet{2016ApJ...824L..23B} further constrained P9 to be of 5--20\,M$_\oplus$ with an eccentricity of $\sim 0.2$--0.8, semi-major axis of $\sim 500$--1050\,AU (perihelion distance of $\sim150$--350\,AU) and inclination about 30$\degr$.
The semi-major axes and eccentricities constrained for P9 are depicted in Fig.~\ref{fig:pz} and they overlap with the region of the outer edges of the Solar system's parking zone.
This means that there was at least one encounter along the solar orbit that could have changed the aphelion velocity of P9 by 100\%.
For such perturbation of P9, the encounter needs to have the appropriate geometry (where the encounter and P9's orbit are in the same plane).
The outer limit of the parking zone serves as an estimate of the level to which a population of bodies orbiting in the System system was perturbed;
that is, the concept of the parking zone assumes that there will be bodies on orbits with certain geometry with respect to the encounter plane. As a consequence, if a single body is orbiting at, or is close to the parking zone (such as P9), the probability of a perturbation occurring  is given by the  probability  to obtain the appropriate geometry of the stellar encounter.  

In this context, \citet{2016ApJ...823L...3L} estimated the probability for the ejection of P9 from its current orbit by field stars.  
Using a large ensamble of simulations with Monte Carlo sampling, they first calculated the cross section for the ejection and then integrated these along the solar orbit, assuming a constant number stellar density of 0.1\,pc$^{-3}$, and a  velocity dispersion of 40\,km\,s$^{-1}$ for 4.6\,Gyr.
They estimate the probability of ejecting P9 due to a passing field star to be $\lesssim3$\%.
Note that while \citet{2016ApJ...823L...3L} considered an isotropic distribution of the direction of the encounters approach, \citet{2014MNRAS.442.3653F} find the distribution non-isotropic (encounters in the direction of the solar antapex are more common).

The existence of P9 is important to establish the existence of the parking zone of the Solar system. In Fig. \ref{fig:pz} we show that the inner edge of the parking zone is delimited by Neptune's perturbing distance (dashed black line). If Planet 9 really exists, the inner edge of the parking zone would be now delimited by its orbital parameters. This means that the inner edge of the parking zone would be shifted towards larger semi-major axes, at $\sim 10^3$~AU. In this case, the Solar system's parking zone would not exist. 

\subsection{Limitations in the computation of stellar encounters}
\label{sect:limitations}

We computed the Galactic stellar encounters in a more complete fashion than in \cite{portegies15}. However, we notice that  our approach has limitations. First, we use different Galaxy models to compute the  local stellar density and the velocity dispersion along the orbit of the Sun. This is inconsistent, because the local density and the stellar velocity dispersion might be different in the two Galaxy models used, even when these models might reproduce the observed properties of the Milky Way locally.  Second, since we used only one snapshot from the N-body Galaxy model, the velocity dispersion along the orbit of the Sun does not evolve with time.

The estimate of the stellar encounters can be improved by computing in a consistent manner the local stellar density and  the velocity dispersion along the orbit of the Sun. This can be achieved by using either the analytical or the N-body Galaxy model. In the analytical Galaxy model, the velocity dispersion can be derived by solving the Jeans equations. In this way the temporal evolution of the velocity dispersion along the orbit of the Sun is also taken into account. However, several assumptions have to be made in order to obtain an uncomplicated solution for $\nu(t, x,y,z)$. For instance, it is necessary to assume an initial velocity dispersion profile  and the velocity ellipsoid aligned with the $R$ and $z$ axes. \citep[][Sect. 2.3]{monari13}. 

In the N-body Galaxy model on the other hand, it is necessary to integrate the orbit of the Sun and to compute $\rho(t,x,y,z)$ and $\nu(t,x,y,z)$ using this model to make a consistent determination of $n_\mathrm{enc}$. To account for the temporal evolution of $\rho$ and $\nu$,  such calculations must include enough snapshots obtained from the N-body simulation. This procedure however,  is not easy to execute given the complexity at handling the huge amount of data  provided by each snapshot in the simulation.   

The improvements mentioned above require further work and  are outside  the scope of this paper.  The computation of the encounter probability by using either of the two methods is left for a future work.


\section{Summary and conclusions}
\label{sect:summary}

We estimate the number of Galactic stellar encounters the Sun may have experienced in the past, along its orbit through the Galaxy.  We aim to improve the previous estimates of the outer edge of the Solar system's parking zone made by \cite{portegies15}. The parking zone is the region in the plane of the eccentricity and  semi-major axis where objects orbiting the Sun have been perturbed by stars belonging to the Sun's birth cluster but not by the planets or by Galactic perturbations.  As a consequence, the orbits of objects located in the parking zone maintain a record of the interaction of the Solar system  with the so called solar siblings \citep{portegies09}.  These orbits carry information that can constrain the natal environment of the Sun. 

We investigate the orbital history of the Sun by using an analytical potential containing a bar and spiral arms  to model the Galaxy. In this potential we integrate the orbit  of the Sun back in time during $4.6$~Gyr.  Since we include the uncertainties in the present-day phase-space coordinates of the Sun, we obtain a collection of possible  orbital histories. Here we study three different orbits, depending on the migration experienced by the Sun namely:  migration inwards, no migration and migration outwards. The Galactic stellar encounters are estimated for  each of these orbits.

We compute the number of stellar encounters ($n_\mathrm{enc}$) by calculating the  frequency of stellar passages experienced by the Sun along its orbit. This frequency is determined by computing the number density and the stellar velocity dispersion along the orbit of the Sun.  We found that $n_\mathrm{enc}= 9.3\times10^4, 28.2\times10^4$ and $17.5\times10^4$ for the  orbits with inward migration, outward migration and no migration respectively. We use these estimates to generate a sample of $n_\mathrm{enc}$ random stellar encounters with certain time of occurrence (\tenc); mass (\menc); pericenter distance (\renc) and velocity (\venc).  By looking at the distribution of stellar encounters in the space of \menc, \venc\ and  \renc , we found that most of the stellar encounters experienced by the Sun have been with low-mass stars (\menc $< 1$~M$_\odot$) with velocities of $20$-$100$~kms$^{-1}$.

We calculate the the outer edge of the Solar system's parking zone using the impulse approximation \citep{rickman76}. For each solar orbit, we calculate the outer edge for 1000 different sets of encounters. The actual outer edge of the Solar system's parking zone is determined  such that the number of encounters along the orbit resulting in smaller  $a_{\mathrm{PZ}}(e)$ is $n_\mathrm{enc}=1$.
The parking zone is then located at about 250--700, 450--950, and 600--1300\,AU (Fig.~\ref{fig:pz}) for the orbits with migration outwards, no migration, and migration inwards, respectively.

Therefore, the orbital history of the Sun is important to establish the outer edge of the parking zone. From Fig.~\ref{fig:pz} it is also clear that the Sun has experienced stronger stellar encounters than those with the Scholz's star.  As a consequence, the location of the outer edge of the parking zone is closer to the Sun than the previous estimates made by  \cite{portegies15} and is comparable to the border between the inner and outer Oort cloud. Regardless of the migration of the solar orbit, we find that objects in the Solar system with semi-major axis smaller than about 200~AU have not been perturbed by encounters with field stars.  However, depending on the migration of the solar orbit, it is possible that the inner Oort cloud (including Sedna) has been perturbed.

We further discuss the effect of the stellar encounters on the stability of the orbit of a hypothetical Planet 9 (P9). According to the orbital parameters of P9, this object is located in the same region as the outer edge of the parking zone. This means that there was at least one encounter along the solar orbit that could have changed the aphelion velocity of P9 by $100\%$.  



\section*{Acknowledgements}
We thank Joris Hense and Inti Pelupessy for helpful discussions.
We thank the reviewer for pointing out several drawbacks in our original methods and for comments that lead to substantial improvement of the presented work.
This work was supported by the Nederlandse Onderzoekschool voor Astronomie (NOVA), the Netherlands Research Council NWO (grants \#639.073.803 [VICI],  \#614.061.608 [AMUSE] and \#612.071.305 [LGM]).

\bibliographystyle{mn2e}
\bibliography{references}

\begin{thebibliography}{}
\makeatletter
\def\mn@urlcharsother{%
\let\do\@makeother
\do\$\do\&\do\#\do\^\do\_\do\%\do\~}
\def\mn@doi{\begingroup
\mn@urlcharsother
\@ifnextchar[%
{\mn@doi@}
{\mn@doi@[]}}
\def\mn@doi@[#1]#2{%
\def\@tempa{#1}%
\ifx\@tempa\@empty
\href{http://dx.doi.org/#2}{doiXX:#2}%
\else
\href{http://dx.doi.org/#2}{#1}%
\fi
\endgroup
}
\def\mn@eprint#1#2{%
\mn@eprint@#1:#2::\@nil}
\def\mn@eprint@arXiv#1{\href{http://arxiv.org/abs/#1}{{\tt arXiv:#1}}}
\def\mn@eprint@dblp#1{\href{http://dblp.uni-trier.de/rec/bibtex/#1.xml}{dblp:#1}}
\def\mn@eprint@#1:#2:#3:#4\@nil{%
\def\@tempa{#1}%
\def\@tempb{#2}%
\def\@tempc{#3}%
\ifx\@tempc\@empty
\let\@tempc\@tempb
\let\@tempb\@tempa
\fi
\ifx\@tempb\@empty
\def\@tempb{arXiv}%
\fi
\@ifundefined{mn@eprint@\@tempb}
{\@tempb:\@tempc}
{\expandafter\expandafter\csname
  mn@eprint@\@tempb\endcsname\expandafter{\@tempc}}%
}

\bibitem[\protect\citeauthoryear{{Allen}}{{Allen}}{1973}]{1973asqu.book.....A}
{Allen} C.~W.,  1973, {Astrophysical quantities}

\bibitem[\protect\citeauthoryear{{Allen} \& {Santill\'an}}{{Allen} \&
  {Santill\'an}}{1991}]{allen}
{Allen} C.,  {Santill\'an} A.,  1991, Rev. Mex. Astron. Astrofis., \href
  {http://adsabs.harvard.edu/abs/1991RMxAA..22..255A} {22, 255}

\bibitem[\protect\citeauthoryear{{Allen}, {Moreno} \& {Pichardo}}{{Allen}
  et~al.}{2006}]{allen06}
{Allen} C.,  {Moreno} E.,    {Pichardo} B.,  2006, \mn@doi [ApJ]
  {10.1086/508676}, \href {http://adsabs.harvard.edu/abs/2006ApJ...652.1150A}
  {652, 1150}

\bibitem[\protect\citeauthoryear{{Antoja}, {Valenzuela}, {Pichardo}, {Moreno},
  {Figueras} \& {Fern{\'a}ndez}}{{Antoja} et~al.}{2009}]{antoja09}
{Antoja} T.,  {Valenzuela} O.,  {Pichardo} B.,  {Moreno} E.,  {Figueras} F.,
  {Fern{\'a}ndez} D.,  2009, \mn@doi [ApJ] {10.1088/0004-637X/700/2/L78}, \href
  {http://adsabs.harvard.edu/abs/2009ApJ...700L..78A} {700, L78}

\bibitem[\protect\citeauthoryear{{Antoja}, {Figueras}, {Romero-G{\'o}mez},
  {Pichardo}, {Valenzuela} \& {Moreno}}{{Antoja} et~al.}{2011}]{antoja}
{Antoja} T.,  {Figueras} F.,  {Romero-G{\'o}mez} M.,  {Pichardo} B.,
  {Valenzuela} O.,    {Moreno} E.,  2011, \mn@doi [MNRAS]
  {10.1111/j.1365-2966.2011.19190.x}, \href
  {http://adsabs.harvard.edu/abs/2011MNRAS.418.1423A} {418, 1423}

\bibitem[\protect\citeauthoryear{{Bailer-Jones}}{{Bailer-Jones}}{2015}]{bailer15}
{Bailer-Jones} C.~A.~L.,  2015, \mn@doi [A\&A] {10.1051/0004-6361/201425221},
  \href {http://adsabs.harvard.edu/abs/2015A26A...575A..35B} {575, A35}

\bibitem[\protect\citeauthoryear{{Batygin} \& {Brown}}{{Batygin} \&
  {Brown}}{2016}]{BB16}
{Batygin} K.,  {Brown} M.~E.,  2016, \mn@doi [AJ] {10.3847/0004-6256/151/2/22},
  \href {http://adsabs.harvard.edu/abs/2016AJ....151...22B} {151, 22}

\bibitem[\protect\citeauthoryear{{B{\'e}dorf}, {Gaburov}, {Fujii}, {Nitadori},
  {Ishiyama} \& {Portegies Zwart}}{{B{\'e}dorf} et~al.}{2014}]{bedorf}
{B{\'e}dorf} J.,  {Gaburov} E.,  {Fujii} M.~S.,  {Nitadori} K.,  {Ishiyama} T.,
     {Portegies Zwart} S.,  2014, in Proceedings of the International
  Conference for High Performance Computing, Networking, Storage and Analysis,
  p. 54-65. pp 54--65, \mn@eprint {arXiv} {1412.0659}

\bibitem[\protect\citeauthoryear{{Bellini}, {Bedin}, {Pichardo}, {Moreno},
  {Allen}, {Piotto} \& {Anderson}}{{Bellini} et~al.}{2010}]{bellini10}
{Bellini} A.,  {Bedin} L.~R.,  {Pichardo} B.,  {Moreno} E.,  {Allen} C.,
  {Piotto} G.,    {Anderson} J.,  2010, \mn@doi [A\&A]
  {10.1051/0004-6361/200913882}, \href
  {http://adsabs.harvard.edu/abs/2010A26A...513A..51B} {513, A51}

\bibitem[\protect\citeauthoryear{{Brown} \& {Batygin}}{{Brown} \&
  {Batygin}}{2016}]{2016ApJ...824L..23B}
{Brown} M.~E.,  {Batygin} K.,  2016, \mn@doi [ApJ]
  {10.3847/2041-8205/824/2/L23}, \href
  {http://adsabs.harvard.edu/abs/2016ApJ...824L..23B} {824, L23}

\bibitem[\protect\citeauthoryear{{Brown}, {Trujillo} \& {Rabinowitz}}{{Brown}
  et~al.}{2004}]{brown04}
{Brown} M.~E.,  {Trujillo} C.,    {Rabinowitz} D.,  2004, \mn@doi [ApJ]
  {10.1086/422095}, \href {http://adsabs.harvard.edu/abs/2004ApJ...617..645B}
  {617, 645}

\bibitem[\protect\citeauthoryear{{Brunini} \& {Fernandez}}{{Brunini} \&
  {Fernandez}}{1996}]{1996A&A...308..988B}
{Brunini} A.,  {Fernandez} J.~A.,  1996, A\&A, \href
  {http://adsabs.harvard.edu/abs/1996A%26A...308..988B} {308, 988}

\bibitem[\protect\citeauthoryear{{Cox} \& {G{\'o}mez}}{{Cox} \&
  {G{\'o}mez}}{2002}]{cox02}
{Cox} D.~P.,  {G{\'o}mez} G.~C.,  2002, \mn@doi [ApJS] {10.1086/341946}, \href
  {http://adsabs.harvard.edu/abs/2002ApJS..142..261C} {142, 261}

\bibitem[\protect\citeauthoryear{{Dones}, {Brasser}, {Kaib} \&
  {Rickman}}{{Dones} et~al.}{2015}]{2015SSRv..197..191D}
{Dones} L.,  {Brasser} R.,  {Kaib} N.,    {Rickman} H.,  2015, \mn@doi [Space
  Sci. Rev.] {10.1007/s11214-015-0223-2}, \href
  {http://adsabs.harvard.edu/abs/2015SSRv..197..191D} {197, 191}

\bibitem[\protect\citeauthoryear{{Drimmel}}{{Drimmel}}{2000}]{drimmel00}
{Drimmel} R.,  2000, A\&A, \href
  {http://adsabs.harvard.edu/abs/2000A\%26A...358L..13D} {358, L13}

\bibitem[\protect\citeauthoryear{{Dybczynski}}{{Dybczynski}}{1994}]{1994CeMDA..58..139D}
{Dybczynski} P.~A.,  1994, \mn@doi [Celestial Mechanics and Dynamical
  Astronomy] {10.1007/BF00695789}, \href
  {http://adsabs.harvard.edu/abs/1994CeMDA..58..139D} {58, 139}

\bibitem[\protect\citeauthoryear{{Dybczy{\'n}ski} \& {Berski}}{{Dybczy{\'n}ski}
  \& {Berski}}{2015}]{dubinsky15}
{Dybczy{\'n}ski} P.~A.,  {Berski} F.,  2015, \mn@doi [MNRAS]
  {10.1093/mnras/stv367}, \href
  {http://adsabs.harvard.edu/abs/2015MNRAS.449.2459D} {449, 2459}

\bibitem[\protect\citeauthoryear{{Feng} \& {Bailer-Jones}}{{Feng} \&
  {Bailer-Jones}}{2014}]{2014MNRAS.442.3653F}
{Feng} F.,  {Bailer-Jones} C.~A.~L.,  2014, \mn@doi [MNRAS]
  {10.1093/mnras/stu1128}, \href
  {http://adsabs.harvard.edu/abs/2014MNRAS.442.3653F} {442, 3653}

\bibitem[\protect\citeauthoryear{{Feng} \& {Bailer-Jones}}{{Feng} \&
  {Bailer-Jones}}{2015}]{2015MNRAS.454.3267F}
{Feng} F.,  {Bailer-Jones} C.~A.~L.,  2015, \mn@doi [MNRAS]
  {10.1093/mnras/stv2222}, \href
  {http://cdsads.u-strasbg.fr/abs/2015MNRAS.454.3267F} {454, 3267}

\bibitem[\protect\citeauthoryear{{Ferrers}}{{Ferrers}}{1877}]{ferrers}
{Ferrers} N.~M.,  1877, Pure Appl. Math., 14, 1

\bibitem[\protect\citeauthoryear{{Fouchard}, {Froeschl{\'e}}, {Rickman} \&
  {Valsecchi}}{{Fouchard} et~al.}{2011}]{2011Icar..214..334F}
{Fouchard} M.,  {Froeschl{\'e}} C.,  {Rickman} H.,    {Valsecchi} G.~B.,  2011,
  \mn@doi [Icarus] {10.1016/j.icarus.2011.04.012}, \href
  {http://adsabs.harvard.edu/abs/2011Icar..214..334F} {214, 334}

\bibitem[\protect\citeauthoryear{{Garc{\'{\i}}a-S{\'a}nchez}, {Weissman},
  {Preston}, {Jones}, {Lestrade}, {Latham}, {Stefanik} \&
  {Paredes}}{{Garc{\'{\i}}a-S{\'a}nchez} et~al.}{2001}]{garcia01}
{Garc{\'{\i}}a-S{\'a}nchez} J.,  {Weissman} P.~R.,  {Preston} R.~A.,  {Jones}
  D.~L.,  {Lestrade} J.-F.,  {Latham} D.~W.,  {Stefanik} R.~P.,    {Paredes}
  J.~M.,  2001, \mn@doi [A\&A] {10.1051/0004-6361:20011330}, \href
  {http://adsabs.harvard.edu/abs/2001A26A...379..634G} {379, 634}

\bibitem[\protect\citeauthoryear{{Gerhard}}{{Gerhard}}{2011}]{gerhard}
{Gerhard} O.,  2011, Memorie della Societa Astronomica Italiana Supplementi,
  \href {http://adsabs.harvard.edu/abs/2011MSAIS..18..185G} {18, 185}

\bibitem[\protect\citeauthoryear{{Gomes}, {Matese} \& {Lissauer}}{{Gomes}
  et~al.}{2006}]{2006Icar..184..589G}
{Gomes} R.~S.,  {Matese} J.~J.,    {Lissauer} J.~J.,  2006, \mn@doi [Icarus]
  {10.1016/j.icarus.2006.05.026}, \href
  {http://adsabs.harvard.edu/abs/2006Icar..184..589G} {184, 589}

\bibitem[\protect\citeauthoryear{{Gomes}, {Soares} \& {Brasser}}{{Gomes}
  et~al.}{2015}]{2015Icar..258...37G}
{Gomes} R.~S.,  {Soares} J.~S.,    {Brasser} R.,  2015, \mn@doi [Icarus]
  {10.1016/j.icarus.2015.06.020}, \href
  {http://adsabs.harvard.edu/abs/2015Icar..258...37G} {258, 37}

\bibitem[\protect\citeauthoryear{{Heisler} \& {Tremaine}}{{Heisler} \&
  {Tremaine}}{1986}]{heisler86}
{Heisler} J.,  {Tremaine} S.,  1986, \mn@doi [Icarus]
  {10.1016/0019-1035(86)90060-6}, \href
  {http://adsabs.harvard.edu/abs/1986Icar...65...13H} {65, 13}

\bibitem[\protect\citeauthoryear{{Holmberg}, {Nordstr{\"o}m} \&
  {Andersen}}{{Holmberg} et~al.}{2009}]{holmberg}
{Holmberg} J.,  {Nordstr{\"o}m} B.,    {Andersen} J.,  2009, \mn@doi [A\&A]
  {10.1051/0004-6361/200811191}, \href
  {http://adsabs.harvard.edu/abs/2009A26A...501..941H} {501, 941}

\bibitem[\protect\citeauthoryear{{Horner} \& {Evans}}{{Horner} \&
  {Evans}}{2002}]{2002MNRAS.335..641H}
{Horner} J.,  {Evans} N.~W.,  2002, \mn@doi [MNRAS]
  {10.1046/j.1365-8711.2002.05649.x}, \href
  {http://adsabs.harvard.edu/abs/2002MNRAS.335..641H} {335, 641}

\bibitem[\protect\citeauthoryear{{Hut} \& {Tremaine}}{{Hut} \&
  {Tremaine}}{1985}]{1985AJ.....90.1548H}
{Hut} P.,  {Tremaine} S.,  1985, \mn@doi [AJ] {10.1086/113868}, \href
  {http://adsabs.harvard.edu/abs/1985AJ.....90.1548H} {90, 1548}

\bibitem[\protect\citeauthoryear{{Jakub{\'{\i}}k} \& {Neslu{\v
  s}an}}{{Jakub{\'{\i}}k} \& {Neslu{\v s}an}}{2008}]{2008CoSka..38...33J}
{Jakub{\'{\i}}k} M.,  {Neslu{\v s}an} L.,  2008, Contributions of the
  Astronomical Observatory Skalnate Pleso, \href
  {http://adsabs.harvard.edu/abs/2008CoSka..38...33J} {38, 33}

\bibitem[\protect\citeauthoryear{{Jakub{\'{\i}}k} \& {Neslu{\v
  s}an}}{{Jakub{\'{\i}}k} \& {Neslu{\v s}an}}{2009}]{2009CoSka..39...85J}
{Jakub{\'{\i}}k} M.,  {Neslu{\v s}an} L.,  2009, Contributions of the
  Astronomical Observatory Skalnate Pleso, \href
  {http://adsabs.harvard.edu/abs/2009CoSka..39...85J} {39, 85}

\bibitem[\protect\citeauthoryear{{J{\'{\i}}lkov{\'a}}, {Carraro}, {Jungwiert}
  \& {Minchev}}{{J{\'{\i}}lkov{\'a}} et~al.}{2012}]{jilkova12}
{J{\'{\i}}lkov{\'a}} L.,  {Carraro} G.,  {Jungwiert} B.,    {Minchev} I.,
  2012, \mn@doi [A\&A] {10.1051/0004-6361/201117347}, \href
  {http://adsabs.harvard.edu/abs/2012A26A...541A..64J} {541, A64}

\bibitem[\protect\citeauthoryear{{J{\'{\i}}lkov{\'a}}, {Portegies Zwart},
  {Pijloo} \& {Hammer}}{{J{\'{\i}}lkov{\'a}} et~al.}{2015}]{jilkova15}
{J{\'{\i}}lkov{\'a}} L.,  {Portegies Zwart} S.,  {Pijloo} T.,    {Hammer} M.,
  2015, \mn@doi [MNRAS] {10.1093/mnras/stv1803}, \href
  {http://adsabs.harvard.edu/abs/2015MNRAS.453.3157J} {453, 3157}

\bibitem[\protect\citeauthoryear{{Juri{\'c}} et~al.,}{{Juri{\'c}}
  et~al.}{2008}]{juric08}
{Juri{\'c}} M.,  et~al., 2008, \mn@doi [ApJ] {10.1086/523619}, \href
  {http://adsabs.harvard.edu/abs/2008ApJ...673..864J} {673, 864}

\bibitem[\protect\citeauthoryear{{Kaib}, {Ro{\v s}kar} \& {Quinn}}{{Kaib}
  et~al.}{2011}]{kaib11}
{Kaib} N.~A.,  {Ro{\v s}kar} R.,    {Quinn} T.,  2011, \mn@doi [Icarus]
  {10.1016/j.icarus.2011.07.037}, \href
  {http://adsabs.harvard.edu/abs/2011Icar..215..491K} {215, 491}

\bibitem[\protect\citeauthoryear{{Kepler}, {Kleinman}, {Nitta}, {Koester},
  {Castanheira}, {Giovannini}, {Costa} \& {Althaus}}{{Kepler}
  et~al.}{2007}]{2007MNRAS.375.1315K}
{Kepler} S.~O.,  {Kleinman} S.~J.,  {Nitta} A.,  {Koester} D.,  {Castanheira}
  B.~G.,  {Giovannini} O.,  {Costa} A.~F.~M.,    {Althaus} L.,  2007, \mn@doi
  [MNRAS] {10.1111/j.1365-2966.2006.11388.x}, \href
  {http://adsabs.harvard.edu/abs/2007MNRAS.375.1315K} {375, 1315}

\bibitem[\protect\citeauthoryear{{Kroupa}}{{Kroupa}}{2001}]{kroupa}
{Kroupa} P.,  2001, \mn@doi [MNRAS] {10.1046/j.1365-8711.2001.04022.x}, \href
  {http://adsabs.harvard.edu/abs/2001MNRAS.322..231K} {322, 231}

\bibitem[\protect\citeauthoryear{{Li} \& {Adams}}{{Li} \&
  {Adams}}{2016}]{2016ApJ...823L...3L}
{Li} G.,  {Adams} F.~C.,  2016, \mn@doi [ApJ] {10.3847/2041-8205/823/1/L3},
  \href {http://adsabs.harvard.edu/abs/2016ApJ...823L...3L} {823, L3}

\bibitem[\protect\citeauthoryear{{Lykawka} \& {Mukai}}{{Lykawka} \&
  {Mukai}}{2008}]{2008AJ....135.1161L}
{Lykawka} P.~S.,  {Mukai} T.,  2008, \mn@doi [AJ]
  {10.1088/0004-6256/135/4/1161}, \href
  {http://adsabs.harvard.edu/abs/2008AJ....135.1161L} {135, 1161}

\bibitem[\protect\citeauthoryear{{Mamajek}}{{Mamajek}}{2016}]{2016_mamajek}
{Mamajek} E.~E.,  2016, Private communication

\bibitem[\protect\citeauthoryear{{Mamajek}, {Barenfeld}, {Ivanov}, {Kniazev},
  {V{\"a}is{\"a}nen}, {Beletsky} \& {Boffin}}{{Mamajek} et~al.}{2015}]{mamajek}
{Mamajek} E.~E.,  {Barenfeld} S.~A.,  {Ivanov} V.~D.,  {Kniazev} A.~Y.,
  {V{\"a}is{\"a}nen} P.,  {Beletsky} Y.,    {Boffin} H.~M.~J.,  2015, \mn@doi
  [ApJ] {10.1088/2041-8205/800/1/L17}, \href
  {http://adsabs.harvard.edu/abs/2015ApJ...800L..17M} {800, L17}

\bibitem[\protect\citeauthoryear{{Mart{\'{\i}}nez-Barbosa}, {Brown} \&
  {Portegies Zwart}}{{Mart{\'{\i}}nez-Barbosa} et~al.}{2015}]{martinezb14}
{Mart{\'{\i}}nez-Barbosa} C.~A.,  {Brown} A.~G.~A.,    {Portegies Zwart} S.,
  2015, \mn@doi [MNRAS] {10.1093/mnras/stu2094}, \href
  {http://adsabs.harvard.edu/abs/2015MNRAS.446..823M} {446, 823}

\bibitem[\protect\citeauthoryear{{Matese} \& {Whitmire}}{{Matese} \&
  {Whitmire}}{1986}]{1986Icar...65...37M}
{Matese} J.~J.,  {Whitmire} D.~P.,  1986, \mn@doi [Icarus]
  {10.1016/0019-1035(86)90062-X}, \href
  {http://adsabs.harvard.edu/abs/1986Icar...65...37M} {65, 37}

\bibitem[\protect\citeauthoryear{{Melita}, {Williams}, {Collander-Brown} \&
  {Fitzsimmons}}{{Melita} et~al.}{2004}]{2004Icar..171..516M}
{Melita} M.~D.,  {Williams} I.~P.,  {Collander-Brown} S.~J.,    {Fitzsimmons}
  A.,  2004, \mn@doi [Icarus] {10.1016/j.icarus.2004.05.014}, \href
  {http://adsabs.harvard.edu/abs/2004Icar..171..516M} {171, 516}

\bibitem[\protect\citeauthoryear{{Minchev} \& {Famaey}}{{Minchev} \&
  {Famaey}}{2010}]{minchev10}
{Minchev} I.,  {Famaey} B.,  2010, \mn@doi [ApJ] {10.1088/0004-637X/722/1/112},
  \href {http://adsabs.harvard.edu/abs/2010ApJ...722..112M} {722, 112}

\bibitem[\protect\citeauthoryear{{Minchev}, {Chiappini} \& {Martig}}{{Minchev}
  et~al.}{2013}]{minchev13}
{Minchev} I.,  {Chiappini} C.,    {Martig} M.,  2013, \mn@doi [A\&A]
  {10.1051/0004-6361/201220189}, \href
  {http://adsabs.harvard.edu/abs/2013A\%26A...558A...9M} {558, A9}

\bibitem[\protect\citeauthoryear{{Miyamoto} \& {Nagai}}{{Miyamoto} \&
  {Nagai}}{1975}]{miyamoto}
{Miyamoto} M.,  {Nagai} R.,  1975, PASJ, \href
  {http://adsabs.harvard.edu/abs/1975PASJ...27..533M} {27, 533}

\bibitem[\protect\citeauthoryear{{Monari}, {Antoja} \& {Helmi}}{{Monari}
  et~al.}{2013}]{monari13}
{Monari} G.,  {Antoja} T.,    {Helmi} A.,  2013, \href
  {http://adsabs.harvard.edu/abs/2013arXiv1306.2632M} {arXiv:1306.2632}

\bibitem[\protect\citeauthoryear{{Monari}, {Helmi}, {Antoja} \&
  {Steinmetz}}{{Monari} et~al.}{2014}]{monari14}
{Monari} G.,  {Helmi} A.,  {Antoja} T.,    {Steinmetz} M.,  2014, \mn@doi
  [A\&A] {10.1051/0004-6361/201423666}, \href
  {http://adsabs.harvard.edu/abs/2014A%26A...569A..69M} {569, A69}

\bibitem[\protect\citeauthoryear{{Murray}}{{Murray}}{1999}]{1999MNRAS.309...31M}
{Murray} J.~B.,  1999, \mn@doi [MNRAS] {10.1046/j.1365-8711.1999.02806.x},
  \href {http://adsabs.harvard.edu/abs/1999MNRAS.309...31M} {309, 31}

\bibitem[\protect\citeauthoryear{{Nordstr{\"o}m} et~al.,}{{Nordstr{\"o}m}
  et~al.}{2004}]{nordstrom04}
{Nordstr{\"o}m} B.,  et~al., 2004, \mn@doi [A\&A] {10.1051/0004-6361:20035959},
  \href {http://adsabs.harvard.edu/abs/2004A\%26A...418..989N} {418, 989}

\bibitem[\protect\citeauthoryear{{Oort}}{{Oort}}{1950}]{oort}
{Oort} J.~H.,  1950, Bull. Astron. Inst. Netherlands, \href
  {http://adsabs.harvard.edu/abs/1950BAN....11...91O} {11, 91}

\bibitem[\protect\citeauthoryear{{Pecaut} \& {Mamajek}}{{Pecaut} \&
  {Mamajek}}{2013}]{2013ApJS..208....9P}
{Pecaut} M.~J.,  {Mamajek} E.~E.,  2013, \mn@doi [ApJS]
  {10.1088/0067-0049/208/1/9}, \href
  {http://adsabs.harvard.edu/abs/2013ApJS..208....9P} {208, 9}

\bibitem[\protect\citeauthoryear{{Pecaut}, {Mamajek} \& {Bubar}}{{Pecaut}
  et~al.}{2012}]{2012ApJ...746..154P}
{Pecaut} M.~J.,  {Mamajek} E.~E.,    {Bubar} E.~J.,  2012, \mn@doi [ApJ]
  {10.1088/0004-637X/746/2/154}, \href
  {http://adsabs.harvard.edu/abs/2012ApJ...746..154P} {746, 154}

\bibitem[\protect\citeauthoryear{{Pelupessy}, {van Elteren}, {de Vries},
  {McMillan}, {Drost} \& {Portegies Zwart}}{{Pelupessy}
  et~al.}{2013}]{pelupessy13}
{Pelupessy} F.~I.,  {van Elteren} A.,  {de Vries} N.,  {McMillan} S.~L.~W.,
  {Drost} N.,    {Portegies Zwart} S.~F.,  2013, \mn@doi [A\&A]
  {10.1051/0004-6361/201321252}, \href
  {http://adsabs.harvard.edu/abs/2013A\%26A...557A..84P} {557, A84}

\bibitem[\protect\citeauthoryear{{Plummer}}{{Plummer}}{1911}]{plummer}
{Plummer} H.~C.,  1911, MNRAS, \href
  {http://adsabs.harvard.edu/abs/1911MNRAS..71..460P} {71, 460}

\bibitem[\protect\citeauthoryear{{Portegies Zwart}}{{Portegies
  Zwart}}{2009}]{portegies09}
{Portegies Zwart} S.~F.,  2009, \mn@doi [ApJ] {10.1088/0004-637X/696/1/L13},
  \href {http://adsabs.harvard.edu/abs/2009ApJ...696L..13P} {696, L13}

\bibitem[\protect\citeauthoryear{{Portegies Zwart} \&
  {J{\'{\i}}lkov{\'a}}}{{Portegies Zwart} \&
  {J{\'{\i}}lkov{\'a}}}{2015}]{portegies15}
{Portegies Zwart} S.~F.,  {J{\'{\i}}lkov{\'a}} L.,  2015, \mn@doi [MNRAS]
  {10.1093/mnras/stv877}, \href
  {http://adsabs.harvard.edu/abs/2015MNRAS.451..144P} {451, 144}

\bibitem[\protect\citeauthoryear{{Portegies Zwart}, {McMillan}, {van Elteren},
  {Pelupessy} \& {de Vries}}{{Portegies Zwart} et~al.}{2013}]{portegies13}
{Portegies Zwart} S.,  {McMillan} S.~L.~W.,  {van Elteren} E.,  {Pelupessy} I.,
     {de Vries} N.,  2013, \mn@doi [Computer Physics Communications]
  {10.1016/j.cpc.2012.09.024}, \href
  {http://adsabs.harvard.edu/abs/2013CoPhC.183..456P} {183, 456}

\bibitem[\protect\citeauthoryear{{Rickman}}{{Rickman}}{1976}]{rickman76}
{Rickman} H.,  1976, Bulletin of the Astronomical Institutes of Czechoslovakia,
  \href {http://adsabs.harvard.edu/abs/1976BAICz..27...92R} {27, 92}

\bibitem[\protect\citeauthoryear{{Rickman}}{{Rickman}}{2014}]{rickman14}
{Rickman} H.,  2014, \mn@doi [Meteoritics and Planetary Science]
  {10.1111/maps.12080}, \href
  {http://adsabs.harvard.edu/abs/2014M26PS...49....8R} {49, 8}

\bibitem[\protect\citeauthoryear{{Rickman}, {Fouchard}, {Valsecchi} \&
  {Froeschl{\'e}}}{{Rickman} et~al.}{2005}]{2005EM&P...97..411R}
{Rickman} H.,  {Fouchard} M.,  {Valsecchi} G.~B.,    {Froeschl{\'e}} C.,  2005,
  \mn@doi [Earth Moon and Planets] {10.1007/s11038-006-9113-7}, \href
  {http://adsabs.harvard.edu/abs/2005EM%26P...97..411R} {97, 411}

\bibitem[\protect\citeauthoryear{{Rickman}, {Fouchard}, {Froeschl{\'e}} \&
  {Valsecchi}}{{Rickman} et~al.}{2008}]{rickman08}
{Rickman} H.,  {Fouchard} M.,  {Froeschl{\'e}} C.,    {Valsecchi} G.~B.,  2008,
  \mn@doi [Celestial Mechanics and Dynamical Astronomy]
  {10.1007/s10569-008-9140-y}, \href
  {http://adsabs.harvard.edu/abs/2008CeMDA.102..111R} {102, 111}

\bibitem[\protect\citeauthoryear{{Romero-G{\'o}mez}, {Athanassoula}, {Antoja}
  \& {Figueras}}{{Romero-G{\'o}mez} et~al.}{2011}]{merce2}
{Romero-G{\'o}mez} M.,  {Athanassoula} E.,  {Antoja} T.,    {Figueras} F.,
  2011, \mn@doi [MNRAS] {10.1111/j.1365-2966.2011.19569.x}, \href
  {http://adsabs.harvard.edu/abs/2011MNRAS.418.1176R} {418, 1176}

\bibitem[\protect\citeauthoryear{{Ro{\v s}kar}, {Debattista}, {Quinn},
  {Stinson} \& {Wadsley}}{{Ro{\v s}kar} et~al.}{2008}]{roskar}
{Ro{\v s}kar} R.,  {Debattista} V.~P.,  {Quinn} T.~R.,  {Stinson} G.~S.,
  {Wadsley} J.,  2008, \mn@doi [ApJ] {10.1086/592231}, \href
  {http://adsabs.harvard.edu/abs/2008ApJ...684L..79R} {684, L79}

\bibitem[\protect\citeauthoryear{{Sch{\"o}nrich}, {Binney} \&
  {Dehnen}}{{Sch{\"o}nrich} et~al.}{2010}]{schonrich}
{Sch{\"o}nrich} R.,  {Binney} J.,    {Dehnen} W.,  2010, \mn@doi [MNRAS]
  {10.1111/j.1365-2966.2010.16253.x}, \href
  {http://adsabs.harvard.edu/abs/2010MNRAS.403.1829S} {403, 1829}

\bibitem[\protect\citeauthoryear{{Sellwood} \& {Binney}}{{Sellwood} \&
  {Binney}}{2002}]{sellwood}
{Sellwood} J.~A.,  {Binney} J.~J.,  2002, \mn@doi [MNRAS]
  {10.1046/j.1365-8711.2002.05806.x}, \href
  {http://adsabs.harvard.edu/abs/2002MNRAS.336..785S} {336, 785}

\bibitem[\protect\citeauthoryear{{Trujillo} \& {Sheppard}}{{Trujillo} \&
  {Sheppard}}{2014}]{trujillo}
{Trujillo} C.~A.,  {Sheppard} S.~S.,  2014, \mn@doi [Nature]
  {10.1038/nature13156}, \href
  {http://adsabs.harvard.edu/abs/2014Natur.507..471T} {507, 471}

\bibitem[\protect\citeauthoryear{{Whitmire} \& {Matese}}{{Whitmire} \&
  {Matese}}{1985}]{1985Natur.313...36W}
{Whitmire} D.~P.,  {Matese} J.~J.,  1985, \mn@doi [Nature] {10.1038/313036a0},
  \href {http://adsabs.harvard.edu/abs/1985Natur.313...36W} {313, 36}

\bibitem[\protect\citeauthoryear{{Wielen}, {Fuchs} \& {Dettbarn}}{{Wielen}
  et~al.}{1996}]{wielen96}
{Wielen} R.,  {Fuchs} B.,    {Dettbarn} C.,  1996, A\&A, \href
  {http://adsabs.harvard.edu/abs/1996A26A...314..438W} {314, 438}

\makeatother
\end{thebibliography}

\label{lastpage}
\bsp
\end{document}